\pdfoutput=1

\documentclass[11pt,twoside,a4paper,cmspaper,final,collab]{cms-tdr}
\def\svnVersion{334129}\def\cmsCernNoTag{CERN-PH-EP/2013-037}\def\cmsCernDate{\today}\def\cmsMessage{Published in Physical Review D as \href{http://dx.doi.org/10.1103/PhysRevD.93.052007}{\doi{10.1103/PhysRevD.93.052007.}}}

\begin{document}\cmsNoteHeader{TOP-14-023}

\hyphenation{had-ron-i-za-tion}
\hyphenation{cal-or-i-me-ter}
\hyphenation{de-vices}
\RCS$Revision: 331182 $
\RCS$HeadURL: svn+ssh://svn.cern.ch/reps/tdr2/papers/TOP-14-023/trunk/TOP-14-023.tex $
\RCS$Id: TOP-14-023.tex 331182 2016-03-08 00:23:55Z linacre $
\newlength\cmsFigWidth
\ifthenelse{\boolean{cms@external}}{\setlength\cmsFigWidth{0.85\columnwidth}}{\setlength\cmsFigWidth{0.4\textwidth}}
\ifthenelse{\boolean{cms@external}}{\providecommand{\cmsLeft}{Top\xspace}}{\providecommand{\cmsLeft}{Left\xspace}}
\ifthenelse{\boolean{cms@external}}{\providecommand{\cmsRight}{Bottom\xspace}}{\providecommand{\cmsRight}{Right\xspace}}

\ifthenelse{\boolean{cms@external}}{\providecommand{\NA}{\ensuremath{\cdots}\xspace}}{\providecommand{\NA}{\text{---}\xspace}}

\cmsNoteHeader{TOP-14-023}
\title{\texorpdfstring{Measurements of \ttbar\ spin correlations and top quark polarization using dilepton final states in \pp collisions at $\sqrt{s}=8\TeV$}{Measurements of ttbar spin correlations and top quark polarization using dilepton final states in pp collisions at sqrt(s) = 8 TeV}}

\date{\today}

\newcommand{\ttfake}{\ttbar\ (non-dileptonic)}
\newcommand{\wjets}{$\PW+\text{jets}$}
\newcommand{\mtop}{$m_{\cPqt}$}
\newcommand{\ttll}{\ensuremath{\ttbar\to\ell^{+}\ell^{-}}\xspace}
\newcommand{\ttlj}{\ensuremath{\ttbar\to\ell+\text{jets}}\xspace}
\newcommand{\pp}{\ensuremath{\Pp\Pp}\xspace}
\newcommand{\ppbar}{\ensuremath{\Pp\Pap}\xspace}
\newcommand{\eepm}{\ensuremath{\Pep\Pem}\xspace}
\newcommand{\mmpm}{\ensuremath{\Pgmp \Pgmm}\xspace}
\newcommand{\ttpm}{\ensuremath{\Pgt^+ \Pgt^-}\xspace}
\newcommand{\empm}{\ensuremath{\Pe^\pm \Pgm^\mp}\xspace}
\newcommand{\ptttbar}{\ensuremath{{p_{\mathrm{T}}}^{\ttbar}}\xspace}

\providecommand{\eqn}{Eq.}
\providecommand{\fig}{Fig.}
\providecommand{\figs}{Figs.}
\providecommand{\tab}{Table}
\providecommand{\tabs}{Tables}
\providecommand{\secn}{Section}
\providecommand{\secns}{Sections}
\providecommand{\reference}{Ref.}
\providecommand{\references}{Refs.}
\providecommand{\FASTJET}{\textsc{fastjet}\xspace}
\providecommand{\NLO}{NLO\xspace}

\newcommand{\relmu}{\mbox{Re} (\hat{\mu}_{\cPqt})}
\newcommand{\imd}{\mbox{Im} (\hat{d}_{\cPqt})}
\newcommand{\dphi}{$\abs{\Delta \phi_{\ell^+\ell^-}}$}

\newcolumntype{X}[1]{D{,}{\,\pm\,}{#1}}

\abstract{
Measurements of the top quark-antiquark ($\ttbar$)
spin correlations and the top quark polarization
are presented for $\ttbar$ pairs produced in \pp collisions at $\sqrt{s}=8\TeV$.
The data correspond to an integrated luminosity of 19.5\fbinv collected with the CMS detector at the LHC. The measurements are performed using events with two oppositely charged leptons (electrons or muons)
and two or more jets, where at least one of the jets is identified as originating from a bottom quark. The spin correlations and polarization are measured from the angular distributions
of the two selected leptons, both inclusively and differentially, with respect to the invariant mass, rapidity, and transverse momentum of the $\ttbar$ system.
The measurements are unfolded to the parton level and found to be in agreement with predictions of the standard model. A search for new physics in the form of anomalous top quark chromo moments
is performed.
No evidence of new physics is observed, and exclusion limits on the real part of the chromo-magnetic dipole moment and the imaginary part of the chromo-electric dipole moment are evaluated.
}

\hypersetup{%
pdfauthor={CMS Collaboration},%
pdftitle={Measurements of ttbar spin correlations and top quark polarization using dilepton final states in pp collisions at sqrt(s) = 8 TeV},%
pdfsubject={CMS},%
pdfkeywords={CMS, physics, top, spin correlations, polarization, asymmetry}}

\maketitle

\section{Introduction}
\label{sec:intro}

The top quark is the heaviest known elementary particle, with mass $m_{\cPqt} = 172.44 \pm 0.48\GeV$~\cite{CMSNewMass}.
The top quark lifetime has been measured as $3.29^{+0.90}_{-0.63} \times 10^{-25}\unit{s}$~\cite{PhysRevD.85.091104},
shorter than the hadronization timescale $1/\Lambda_{\text{QCD}}\approx 10^{-24}\unit{s}$, where $\Lambda_{\text{QCD}}$ is the
quantum chromodynamics (QCD) scale parameter, and also shorter than the spin decorrelation time scale
$m_{\cPqt}/\Lambda^{2}_{\text{QCD}}\approx 10^{-21}\unit{s}$~\cite{MahlonParke2010}.
Consequently,
measurements of the angular distributions of top quark decay products give
access to the spin of the top quark,
allowing the precise testing of perturbative QCD in the top quark--antiquark pair (\ttbar) production process.

At the CERN LHC, top quarks are produced abundantly, predominantly in pairs.
In the standard model (SM), top quarks from pair production have only a small net polarization arising from
electroweak corrections to the QCD-dominated production process,
but the pairs have significant spin correlations~\cite{Bernreuther2013115}.
For low \ttbar\ invariant masses, the production is dominated by the fusion of pairs of gluons with the same helicities,
resulting in the creation of top quark pairs with antiparallel spins in the \ttbar\ center-of-mass frame.
For larger \ttbar\ invariant masses, the dominant production is
via the fusion of gluons with opposite helicities, resulting in \ttbar\ pairs with parallel spins~\cite{MahlonParke2010}.
For models beyond the SM, couplings of the top quark to new particles can alter both the top quark polarization and
the strength of the spin correlations in the \ttbar\ system~\cite{ref:Krohn,ref:Fajfer,1742-6596-447-1-012015,Bernreuther2013115}.

The charged lepton ($\ell$) from the decay $\cPqt \to \cPqb \PWp \to \cPqb \ell^{+} \cPgn_\ell$ is the best spin analyzer among the top quark decay products~\cite{Brandenburg2002235},
and is sensitive to the top quark spin through the helicity angle $\theta^{\star}_\ell$. This is the angle of the lepton in the rest frame of its parent top quark or antiquark,
measured in the helicity frame (i.e.,\,relative to the direction of the parent quark momentum in the \ttbar\ center-of-mass frame)~\cite{Bernreuther2013115}.

For the decay $\ttbar \to \cPqb \ell^{+} \cPgn_\ell \: \cPaqb \ell^{-} \cPagn_\ell$, the
difference in azimuthal angle of the charged leptons in the laboratory frame, $\Delta \phi_{\ell^+\ell^-}$, is sensitive to \ttbar\ spin correlations
and can be measured precisely without reconstructing the full $\ttbar$ system~\cite{MahlonParke2010}.
With the \ttbar\ system fully reconstructed, the opening angle $\varphi$ between the two lepton momenta measured in the rest frames of their respective parent top quark or antiquark is directly sensitive to spin correlations,
as is the product of the cosines of the helicity angles of the two leptons, $\cos\theta^{\star}_{\ell^+} \cos\theta^{\star}_{\ell^-}$~\cite{Bernreuther2013115}.

Recent spin correlation and polarization measurements from the CDF, D0, and ATLAS Collaborations used template fits to angular distributions, and their results were consistent with the SM expectations~\cite{CDFSpinCor,Abazov:2011gi,D0SpinCor,PhysRevD.87.011103,Aad:1564320,Aad:2014mfk}.
In this analysis, the measurements are made using asymmetries in angular distributions unfolded to the parton level,
allowing direct comparisons between the data and theoretical predictions.  The analysis strategy is similar to that presented in \reference~\cite{Chatrchyan:2013wua}; however the larger data set used here and improvements in
the \ttbar\ system reconstruction techniques lead to a reduced statistical uncertainty in the measurements. Furthermore, an improved unfolding technique allows for differential
measurements, which were not presented in \reference~\cite{Chatrchyan:2013wua}.

The polarization $P^{\pm}$ of the top quark (antiquark) in the helicity basis is given by $P^{\pm} = 2A_{P\pm}$~\cite{Bernreuther2013115}, where the asymmetry variable $A_{P\pm}$ is defined as
\begin{equation*}
A_{P\pm} = \frac{N \left(\cos\theta^{\star}_{\ell^\pm} > 0\right)-N \left(\cos\theta^{\star}_{\ell^\pm} < 0\right)}{N \left(\cos\theta^{\star}_{\ell^\pm} > 0\right)+N \left(\cos\theta^{\star}_{\ell^\pm} < 0\right)},
\end{equation*}
where the numbers of events $N \left(\cos\theta^{\star}_{\ell^\pm} > 0\right)$ and $N \left(\cos\theta^{\star}_{\ell^\pm} < 0\right)$ are counted using the helicity angle of the positively (negatively) charged lepton in each event.
Assuming CP invariance, these two measurements can be combined to give the SM polarization $P = 2A_{P} = \left(A_{P+} + A_{P-}\right)$.
Alternatively, the variable $P^{\mathrm{CPV}} = 2A_{P}^{\mathrm{CPV}} = \left(A_{P+} - A_{P-}\right)$ measures possible polarization introduced by a maximally CP-violating process~\cite{Bernreuther2013115}.

For \ttbar\ spin correlations, the variable
\begin{equation*}
A_{\Delta\phi} = \frac{N (  \abs{\Delta \phi_{\ell^+\ell^-}} > \pi/2 )-N ( \abs{\Delta \phi_{\ell^+\ell^-}} < \pi/2 )}{N ( \abs{\Delta \phi_{\ell^+\ell^-}} > \pi/2 ) +N ( \abs{\Delta \phi_{\ell^+\ell^-}} < \pi/2 )}
\end{equation*}
discriminates between correlated and uncorrelated \cPqt\ and \cPaqt\ spins, while the variable
\begin{equation*}
A_{c_{1}c_{2}} = \frac{N (c_{1} c_{2} > 0)-N (c_{1} c_{2} < 0)}{N (c_{1} c_{2} > 0)+N (c_{1} c_{2} < 0)},
\end{equation*}
where $c_{1} = \cos\theta^{\star}_{\ell^+}$ and $c_{2} = \cos\theta^{\star}_{\ell^-}$,
provides a direct measure of the spin correlation coefficient $C_{\mathrm{hel}}$ through the relationship $C_{\mathrm{hel}} = -4 A_{c_{1}c_{2}}$~\cite{Bernreuther2013115}. The variable
\begin{equation*}
A_{\cos\varphi} = \frac{N \left(\cos\varphi > 0\right)-N \left(\cos\varphi < 0\right)}{N \left(\cos\varphi > 0\right)+N \left(\cos\varphi < 0\right)}
\end{equation*}
provides a direct measure of the spin correlation coefficient $D$ by the relation $D = -2 A_{\cos\varphi}$~\cite{Bernreuther2013115}.

In addition to the inclusive measurements, we determine the asymmetries differentially as a function of
three variables describing the \ttbar\ system in the laboratory frame: its invariant mass $M_{\ttbar}$, rapidity $y_{\ttbar}$, and transverse momentum $\ptttbar$.
The results presented in this paper are based on data collected by the CMS experiment at the LHC,
corresponding to an integrated luminosity of 19.5\fbinv from \pp collisions at $\sqrt{s}=8\TeV$.

\section{The CMS detector}

The central feature of the CMS apparatus is a superconducting solenoid of 6\unit{m} internal diameter, providing a magnetic field of 3.8\unit{T}. Within the solenoid volume are a silicon pixel and strip tracker, a lead tungstate crystal electromagnetic calorimeter, and a brass and scintillator hadron calorimeter, each composed of a barrel and two endcap sections. Forward calorimeters extend the pseudorapidity coverage provided by the barrel and endcap detectors. Muons are measured in gas-ionization detectors embedded in the steel flux-return yoke outside the solenoid. The first level of the CMS trigger system, composed of custom hardware processors, uses information from the calorimeters and muon detectors to select the most interesting events in a fixed time interval of less than 4\mus. The high-level trigger processor farm further decreases the event rate from around 100\unit{kHz} to less than 1\unit{kHz}, before data storage. A more detailed description of the CMS detector, together with a definition of the coordinate system used and the relevant kinematic variables, can be found in \reference~\cite{JINST}.

\section{Event samples}
\label{sec:eventsel}

\subsection{Object definition and event selection}
\label{sec:presel}

Events are selected using triggers that require the presence of at least two leptons (electrons or muons) with transverse
momentum (\pt) greater than $17\GeV$ for the highest-\pt lepton and $8\GeV$ for the second-highest \pt lepton.
The trigger efficiency per lepton, measured relative to the full offline lepton selection detailed in this section using a data sample of Drell--Yan ($\ensuremath{\cPZ/\Pgg^\star} \to \ell \ell$) events, is about 98\%
(96\%) for electrons (muons), with variations at the level of several percent depending on the pseudorapidity $\eta$ and \pt\ of the lepton.

The particle-flow (PF) algorithm~\cite{CMS-PAS-PFT-09-001,CMS-PAS-PFT-10-001} is used to reconstruct and
identify each individual particle with an optimized combination of information from the various elements of the CMS detector.
The offline selection requires events to have exactly two leptons of opposite charge
with $\pt > 20 \GeV$ and $\abs{\eta} < 2.4$.
Electron candidates are reconstructed starting from a cluster of energy deposits in the
electromagnetic calorimeter. The cluster is then matched to
a reconstructed track.
The electron selection is based on the shower shape,
track-cluster matching, and consistency between the cluster energy and
the track momentum~\cite{Khachatryan:2015hwa}.
Muon candidates are reconstructed by performing a global fit that
requires
consistent hit patterns
in the silicon tracker and the muon system~\cite{MUOART}.

The events with an $\Pep\Pem$  or $\Pgmp\Pgmm$ pair having an invariant mass, $M_{\ell\ell}$, within 15 \GeV of the \cPZ~boson mass
are removed to suppress the Drell--Yan background.
For all events, we require $M_{\ell\ell}>20$~\GeV.
Leptons are required to be isolated from other activity in the event.
The lepton isolation is measured using the scalar \pt\ sum ($\pt^\text{sum}$)
of all PF particles not associated with the lepton within a cone of radius $\Delta R \equiv\sqrt{\smash[b]{(\Delta\eta)^2+(\Delta\phi)^2}} = 0.3$,
where $\Delta \eta$ ($\Delta \phi$) is the distance in $\eta$ ($\phi$)
between the directions of the lepton and the PF particle at the primary interaction vertex~\cite{TRK-11-001}.
The average contribution of particles from additional \pp interactions in the same or nearby bunch crossings (pileup)
is estimated and subtracted from the $\pt^\text{sum}$ quantity. The isolation requirement is
$\pt^\text{sum} <  \min(5\GeV,\, 0.15 \, \pt^{\ell})$, where $\pt^{\ell}$ is the lepton \pt. Typical lepton identification and isolation efficiencies, measured in
samples of Drell--Yan events~\cite{SUS-13-011}, are 76\% for electrons and 91\% for muons, with variations at the level of several percent
within the \pt\ and $\eta$ ranges of the selected leptons.

The PF particles are clustered to form jets using the anti-\kt clustering algorithm~\cite{antikt}
with a distance parameter of 0.5, as implemented in the {\FASTJET} package~\cite{Cacciari:2011ma}.
The contribution to the jet energy from pileup is estimated on an event-by-event basis using the
jet-area method described in \reference~\cite{cacciari-2008-659}, and is subtracted from the overall jet \pt.
Jets from pileup interactions are suppressed using a multivariate
discriminant based on the multiplicity of objects clustered in the jet,
the jet shape, and the impact parameters of the charged tracks in the jet
with respect to the primary interaction vertex. The jets must be separated from the
selected leptons by $\Delta  R  >  0.4$.

The selected events are required to contain at least two jets with $\pt > 30\GeV$ and  $\abs{\eta} < 2.4$.
At least one of these jets must be consistent with containing
the decay of a bottom ($\PQb$) flavored hadron, as identified using the medium
operating point of the combined secondary vertex (CSV) $\PQb$~quark tagging algorithm~\cite{ref:btag}. We refer to such jets as $\PQb$-tagged jets.
The efficiency of this algorithm for
$\PQb$~quark jets in the $\pt$ range 30--400\GeV is
60--75\% for $\abs{\eta} < 2.4$. The misidentification rate for light-quark or gluon jets is approximately
1\% for the chosen working point~\cite{ref:btag}.

The missing transverse momentum vector \ptvecmiss is defined as the projection on the plane perpendicular to the beam direction of the negative vector sum of the momenta of all reconstructed particles in the event. Its magnitude is referred to as \ETmiss.
The calibrations that are applied to the energy measurements of jets are
propagated to a correction of \ptvecmiss.
The \MET\ value is required to exceed 40\GeV in events with same-flavor leptons in order to further suppress the Drell--Yan background.
There is no \MET\ requirement for $\Pe^{\pm}\Pgm^{\mp}$ events.

\subsection{Signal and background simulation}
\label{sec:mc}

Simulated signal \ttbar\ events with a top quark mass of $m_{\cPqt}=172.5\GeV$
and with SM spin correlations
are generated using the \MCATNLO3.41~\cite{mc@nlo,MCatNLO2} Monte Carlo (MC) event generator with
the CTEQ6M parton distribution functions (PDF)~\cite{Pumplin:2002vw}.
The parton showering and
fragmentation are performed by \HERWIG6.520~\cite{herwig6}.
Simulations with different values of $m_{\cPqt}$ and renormalization and factorization scales ($\mu_\mathrm{R}$ and $\mu_\mathrm{F}$) are used to evaluate the associated
systematic uncertainties. Background samples of \wjets, Drell--Yan, diboson ($\PW\PW$, $\PW\cPZ$, and $\cPZ\cPZ$), triboson, and $\ttbar+\text{boson}$
events are generated with \MADGRAPH~5.1.3.30~\cite{Alwall:2011uj,MADGRAPHNEW},
and normalized to the calculated next-to-leading-order (NLO)~\cite{xsec_MCFM,Campbell:2011bn,MCatNLO3,xsec_ttbarW,xsec_ttbarZ} or next-to-next-to-leading-order (NNLO)~\cite{xsec_WZ} cross sections.
Single top quark events are generated using \POWHEG1.0~\cite{Nason:2004rx,Frixione:2007vw,Alioli:2009je,Alioli:2010xd,Re:2010bp},
and normalized to the theoretical NNLO cross sections~\cite{Kidonakis:2011wy,Kidonakis:2010tc,Kidonakis:2010ux}.
For the background samples and an alternative \ttbar\ sample generated using \POWHEG1.0, the parton showering and fragmentation are done using \PYTHIA6.4.22~\cite{Pythia}.

For both signal and background events, pileup interactions are simulated with \PYTHIA\ and superimposed on the hard collisions
using a pileup multiplicity distribution that reflects the luminosity profile of the analyzed data.
The CMS detector response is simulated using a \GEANTfour-based model~\cite{Geant}.
The simulated events are reconstructed and analyzed with the same software used to process the collision data.

The measured trigger efficiencies are used to weight the simulated events to account for the trigger requirement.
Small differences between the $\PQb$~tagging efficiencies measured in data and simulation~\cite{ref:btag} are
accounted for by using data-to-simulation correction factors to adjust the $\PQb$~tagging probability in simulated events,
while the lepton selection efficiencies (reconstruction, identification, and isolation) are found to be consistent between data and
simulation~\cite{SUS-13-011}.

\section{Background estimation}
\label{Sec:BkgEst}

Control regions (CR) are used to validate the background estimates from simulation and derive scale factors (SF) and systematic uncertainties for some background processes.
Each SF multiplies the simulated background yield for the given process in the signal region (SR) to obtain the final background prediction. The CRs are designed to have similar kinematics to the SR, but with one or two selection requirements reversed, thus enhancing
different SM contributions. The main CRs used in this analysis and the values of the derived SFs are summarized in \tab~\ref{tab:crdef}.

\begin{table*}[!tphb]
\begin{center}
\topcaption{\label{tab:crdef}
Descriptions of the various control regions, their intended background process, and the scale factors derived from them, including either the statistical and systematic uncertainties or the total uncertainty.
The last row gives the scale factor used for all the remaining backgrounds, whose contributions are estimated from simulation alone.}
\begin{scotch}{l| c  c}
Selection change with &   \multirow{2}{*}{Target background process}  &    \multirow{2}{*}{Scale factor}    \\
respect to the signal region \\
\hline
ee or $\mu\mu$ only, & \multirow{2}{*}{$\ensuremath{\cPZ/\Pgg^\star}(\rightarrow \mathrm{ee}/\mu\mu)+\text{jets}$} & \multirow{2}{*}{1.36 $\pm$ 0.02 (stat) $\pm$ 0.2 (syst)} \\
\multicolumn{1}{c|}{$76<M_{\ell\ell}<106\GeV$}  \\
ee or $\mu\mu$ only, no \MET\ req.,  &  \multirow{2}{*}{$\ensuremath{\cPZ/\Pgg^\star}(\rightarrow \tau\tau)+\text{jets}$}  &  \multirow{2}{*}{1.18 $\pm$ 0.01 (stat) $\pm$ 0.1 (syst)}\\
\multicolumn{1}{c|}{$76<M_{\ell\ell}<106\GeV$} \\
\multirow{2}{*}{Same-charge leptons}
& \multirow{2}{*}{ One-lepton processes}   & \multirow{2}{*}{2.2 $\pm$ 0.3 (stat) $\pm$ 1.0 (syst)} \\
 \\
\multirow{2}{*}{Exactly one jet}
& \multirow{2}{*}{Single top quark ($\PQt\PW$, 2 leptons)} &  \multirow{2}{*}{ 1.00 $\pm$ 0.25 (total) }\\
  \\
\multirow{2}{*}{Simulation}
& \multirow{2}{*}{All other backgrounds} &  \multirow{2}{*}{ 1.0 $\pm$ 0.5 (total) }\\
  \\
\end{scotch}
\end{center}
\end{table*}

\begin{table*}[!hptb]
\centering
\topcaption{\label{tab:yields1}
Predicted background and observed event yields,
with their statistical uncertainties, after applying the event selection
criteria and normalization described in the text.}
\setlength{\extrarowheight}{1.5pt}
\begin{scotch}{l | X{4.4} X{4.4} X{4.4} X{4.4} }
  Sample
& \multicolumn{1}{c}{ee}
& \multicolumn{1}{c}{$\mu\mu$}
& \multicolumn{1}{c}{e$\mu$}
& \multicolumn{1}{c}{Total} \\
\hline
              Single top quark ($\PQt\PW$, 2 leptons)   &     298.0 , 1.6   &     425.9 , 1.9   &    1161.9 , 3.1   &    1885.8 , 4.0  \\
                      Single top quark (other)   &       2.6 , 0.6   &       4.6 , 0.9   &      18.8 , 1.6   &      26.1 , 1.9  \\
                                        \ttlj\   &     107.1 , 7.7   &      62.2 , 5.4   &       327 , 13    &       497 , 16  \\
                                       \wjets\   &      7.3  , 3.6   &       1.8 , 1.8   &      10.0 , 3.5   &      19.1 , 5.3  \\
                 $\ensuremath{\cPZ/\Pgg^\star}(\rightarrow \mathrm{ee}/\mu\mu)+\text{jets}$   &    211 , 16 &      368  , 23    &       1.6 , 0.5   &    581 , 28  \\
                  $\ensuremath{\cPZ/\Pgg^\star}(\rightarrow \tau\tau)+\text{jets}$   &  33.9 , 2.5   &      51.5 , 3.0   &     137.6 , 5.1   &     223.0 , 6.4  \\
             $\PW\PW$/$\PW\PZ$/$\PZ\PZ$   &      27.6 , 1.4   &      40.7 , 1.4   &      89.3 , 2.3   &     157.5 , 3.0  \\
                                      Triboson   &       1.5 , 0.1   &       2.3 , 0.2   &       5.2 , 0.3   &       9.0 , 0.4  \\
                                $\ttbar\PW/\ttbar\PZ/\ttbar\gamma$   &      86.4 , 6.5   &     141.3 , 8.2   &      332 , 13     &    559 , 17  \\
\hline
                           Total background   &    775 , 20   &   1098 , 25   &   2083, 20   &   3957 , 38  \\
\hline
                                          Data
& \multicolumn{1}{c}{7089}
& \multicolumn{1}{c}{10074}
& \multicolumn{1}{c}{26735}
& \multicolumn{1}{c}{43898} \\
\hline
              Signal yield (data $-$ background)   &   6314 , 86 &   8980 , 100 &  24650 , 160 &  39940 , 210 \\
\end{scotch}
\end{table*}

For Drell--Yan events, the SF accounts for mismodeling of the \MET distribution (coming largely from mismeasured jets) and mismodeling
of the heavy-flavor content. Only the latter is relevant for  $ \cPZ/\Pgg^\star (\rightarrow \tau\tau)+\text{jets}$, where the \MET is dominated by the well-modeled
undetected neutrinos, so we omit the \MET mismodeling in the derivation of the SF for this process.
The systematic uncertainties in the SFs are taken from the envelope of the variation observed between the three dilepton flavor combinations and in various
alternative CRs.
The CR for single top quark production in association with a \PW~boson ($\PQt\PW$) is still dominated by signal events (75\%), with only a 16\% contribution from $\PQt\PW$ production, which is an enhancement by a factor of 4 compared to the SR.
Given the good agreement between data and simulation in this CR,
we assume a SF of unity for $\PQt\PW$ production, with an uncertainty
of 25\% based on the recent CMS $\PQt\PW$ cross section measurement of $23.4 \pm 5.4\unit{pb}$~\cite{PhysRevLett.112.231802}.

Contributions to the background from diboson and triboson production, as well as \ttbar\ production in association with a boson,
are estimated from simulation. Recent measurements from the CMS Collaboration~\cite{ref:WWWZ,Chatrchyan:2014aqa,Khachatryan:2015sha} indicate agreement
between the predicted and measured cross sections for these processes, and we assign a systematic uncertainty of 50\%.

\section{Event yields and measurements at the reconstruction level}
\label{sec:yields}

The expected background and observed event yields for different dilepton flavor combinations are listed in \tab~\ref{tab:yields1}.
The total predicted yield in the $\Pe\Pgm$ channel is significantly larger than for the same-flavor channels because of the additional requirements on \MET\ and $M_{\ell\ell}$ described in \secn~\ref{sec:eventsel} that are applied to suppress the Drell--Yan background.
After subtraction of the predicted background yields, the remaining yield in the data is assumed to be a signal from dileptonic \ttbar\ decays,
including $\tau$ leptons that decay leptonically. All other \ttbar\ decay modes are treated as background and are included in the \ttlj\ category.
The largest background comes from $\PQt\PW$ production with dileptonic decays.

While the $\abs{\Delta \phi_{\ell^+\ell^-}}$ measurement relies only on the leptonic information, the measurements based on
$\cos\varphi$ and $\cos\theta^{\star}_{\ell^\pm}$
require the reconstruction of the entire \ttbar\  system.
Each signal event has two neutrinos in the final state, and there is also a twofold ambiguity in combining the $\PQb$~quark jets with the leptons.
In the case of events with only one $\PQb$-tagged jet (62\% of the selected events), the untagged jet with the highest $\PQb$~quark likelihood from the CSV algorithm is assumed to be the second $\PQb$~quark jet.
Analytical solutions for the two neutrino momenta are obtained from the measured \ptvecmiss with constraints on the invariant masses of the top quark and \PW~boson decay systems of $m_{\cPqt}=172.5\GeV$ and $m_{\PW}=80.385\GeV$.
Each event can have up to 8 possible solutions. The one most likely to represent the correct \ttbar\ configuration is chosen
based on the probabilities to observe the extracted Bjorken $x$ values of the initial-state partons and the measured lepton energies in their parent top quark rest frames~\cite{ref:1105.5661}.
For events with no physical solutions, a method is used to find a solution with the vector sum of the \pt\ of the two neutrinos as close as possible to the measured \ptvecmiss~\cite{Dalitz1992225,Betchart2014169}.
Nevertheless, in 16\% of the events, no solutions can be found, both in the data and the simulation. These events are not used, except in the inclusive measurement of $\abs{\Delta \phi_{\ell^+\ell^-}}$.

\begin{figure*}[!htpb]
\centering
\includegraphics[width=0.325\linewidth]{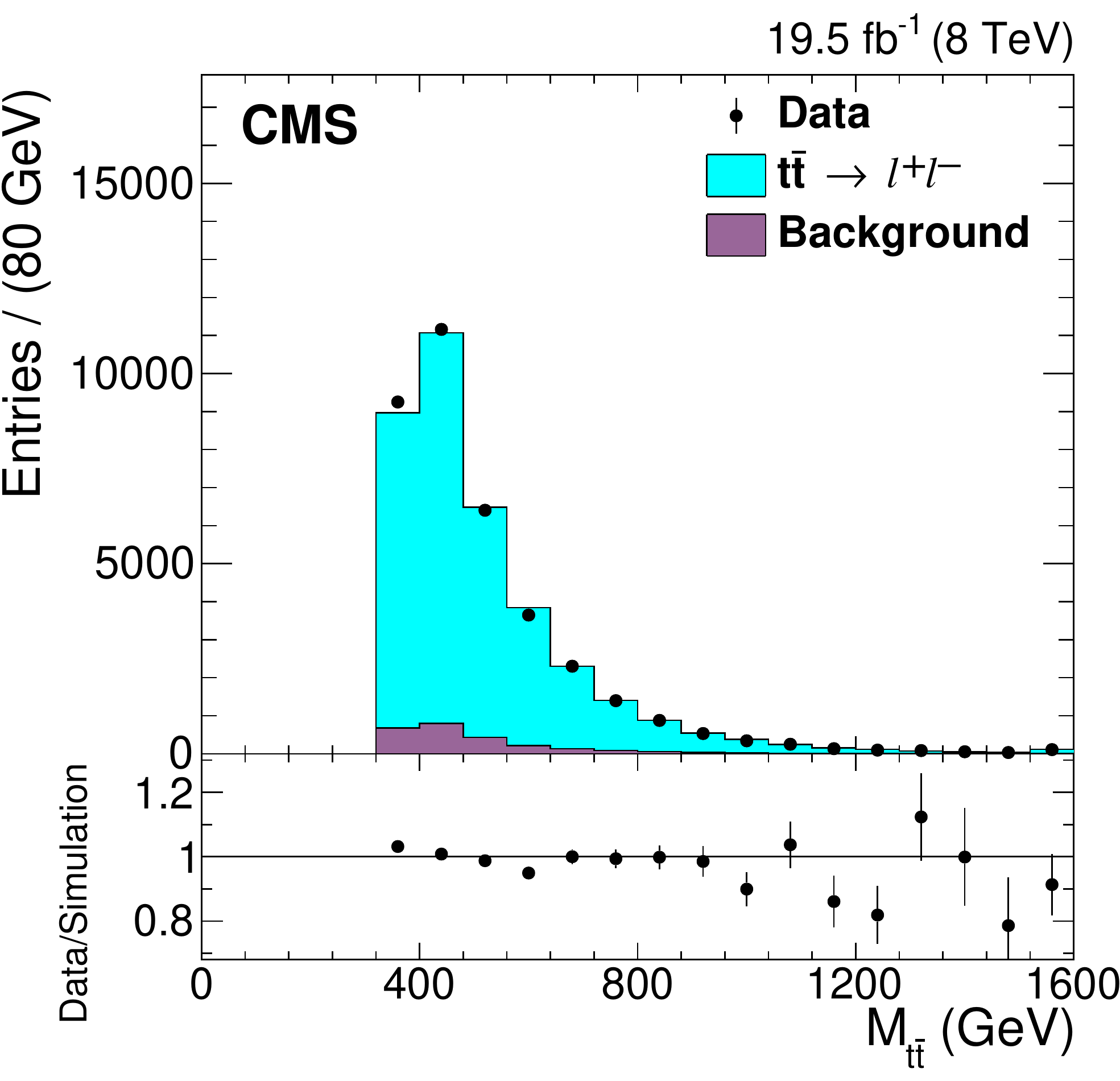}
\includegraphics[width=0.325\linewidth]{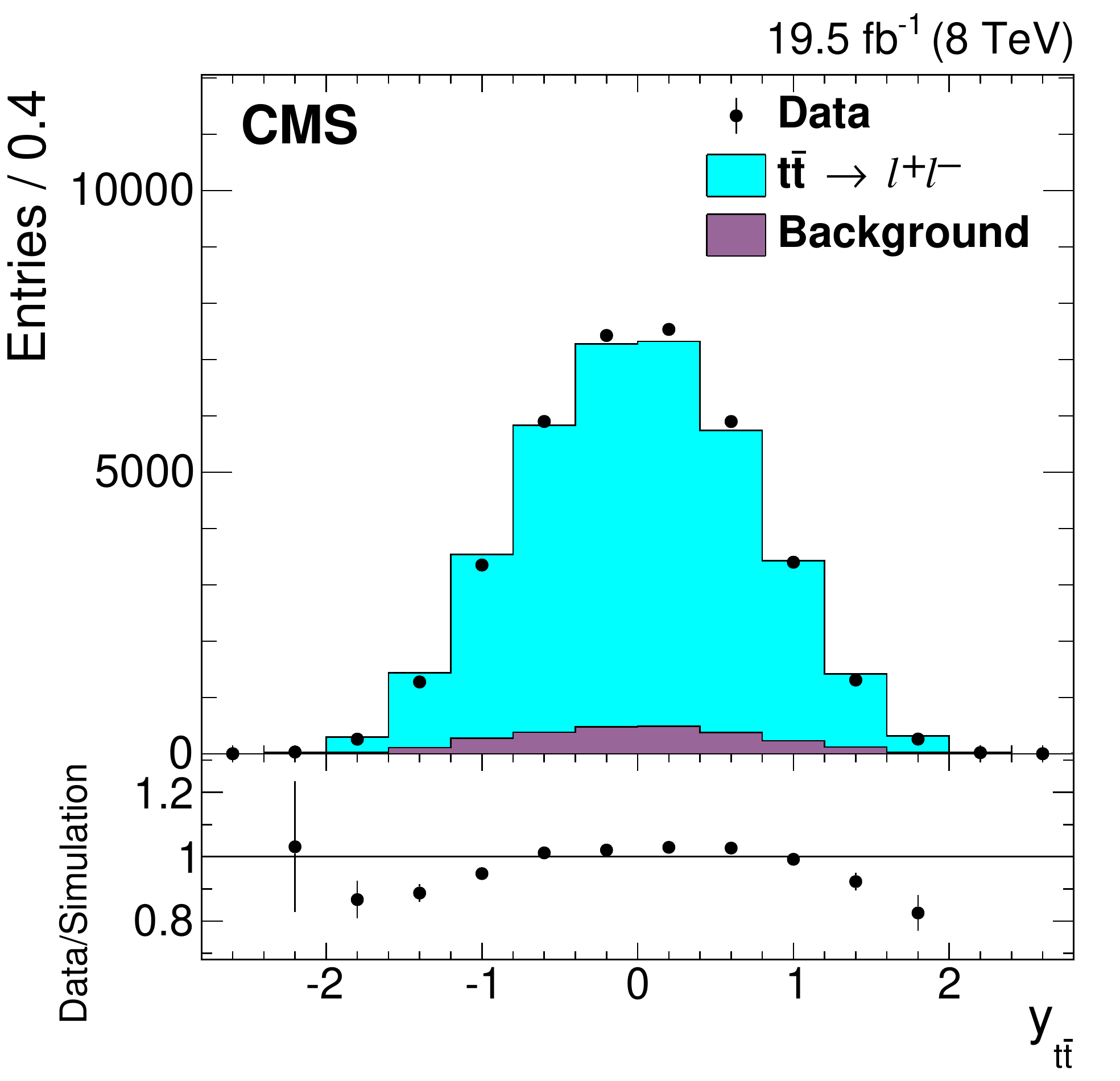}
\includegraphics[width=0.325\linewidth]{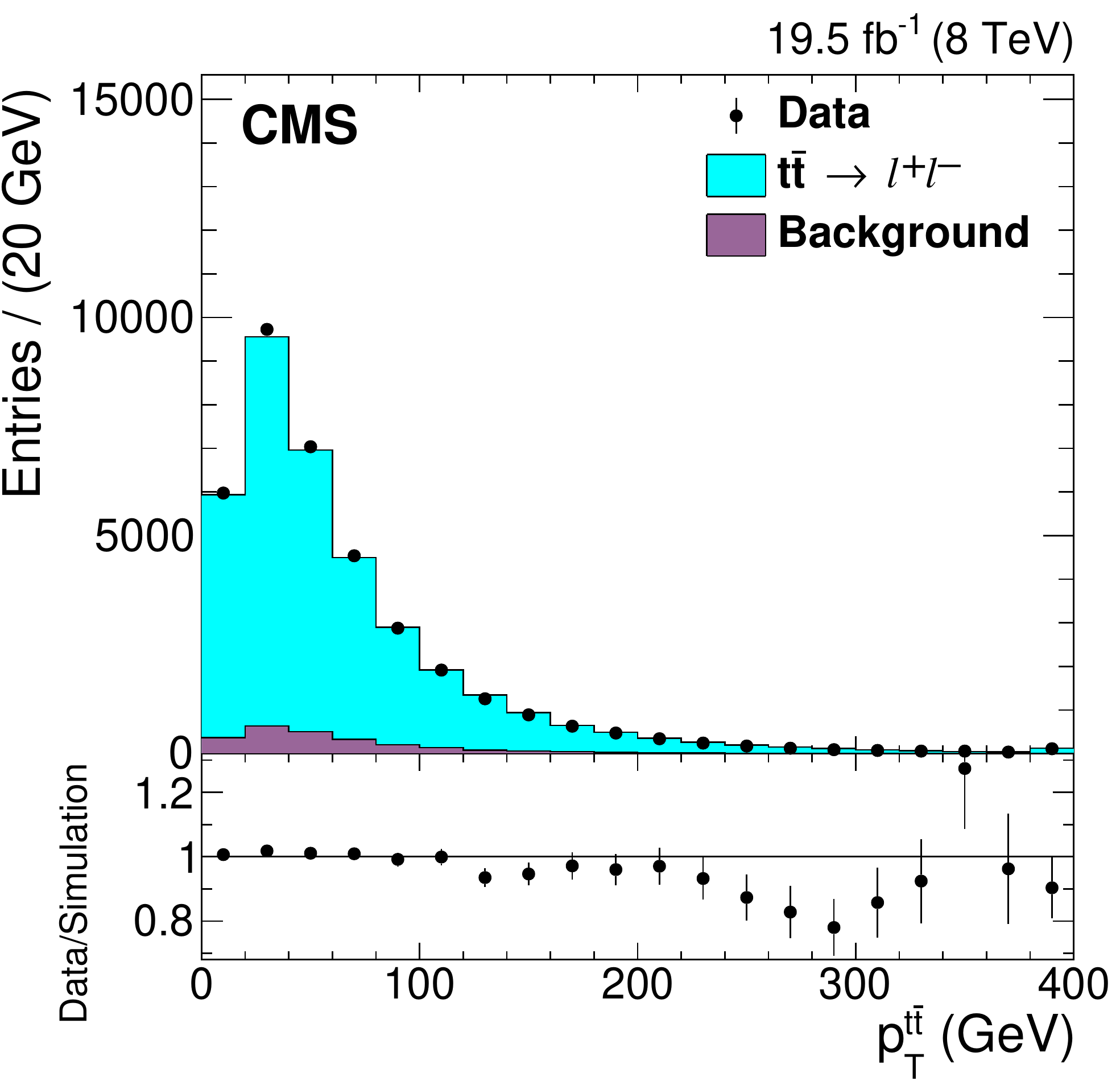}
\caption{\label{fig:preselcomp}\protect
Reconstructed $M_{\ttbar}$, $y_{\ttbar}$, and $\ptttbar$ distributions from data (points) and simulation (histogram), with the expected signal (\ttll) and background distributions shown separately.
All three dilepton flavor combinations are included. The simulated signal yield is normalized to that of the background-subtracted data.
The last bins of the $M_{\ttbar}$ and $\ptttbar$ distributions include overflow events.
The vertical bars on the data points represent the statistical uncertainties.
The lower panels show the ratio of the numbers of events from data and simulation.
}
\end{figure*}

\begin{figure*}[!htpb]
\centering
\includegraphics[width=0.325\linewidth]{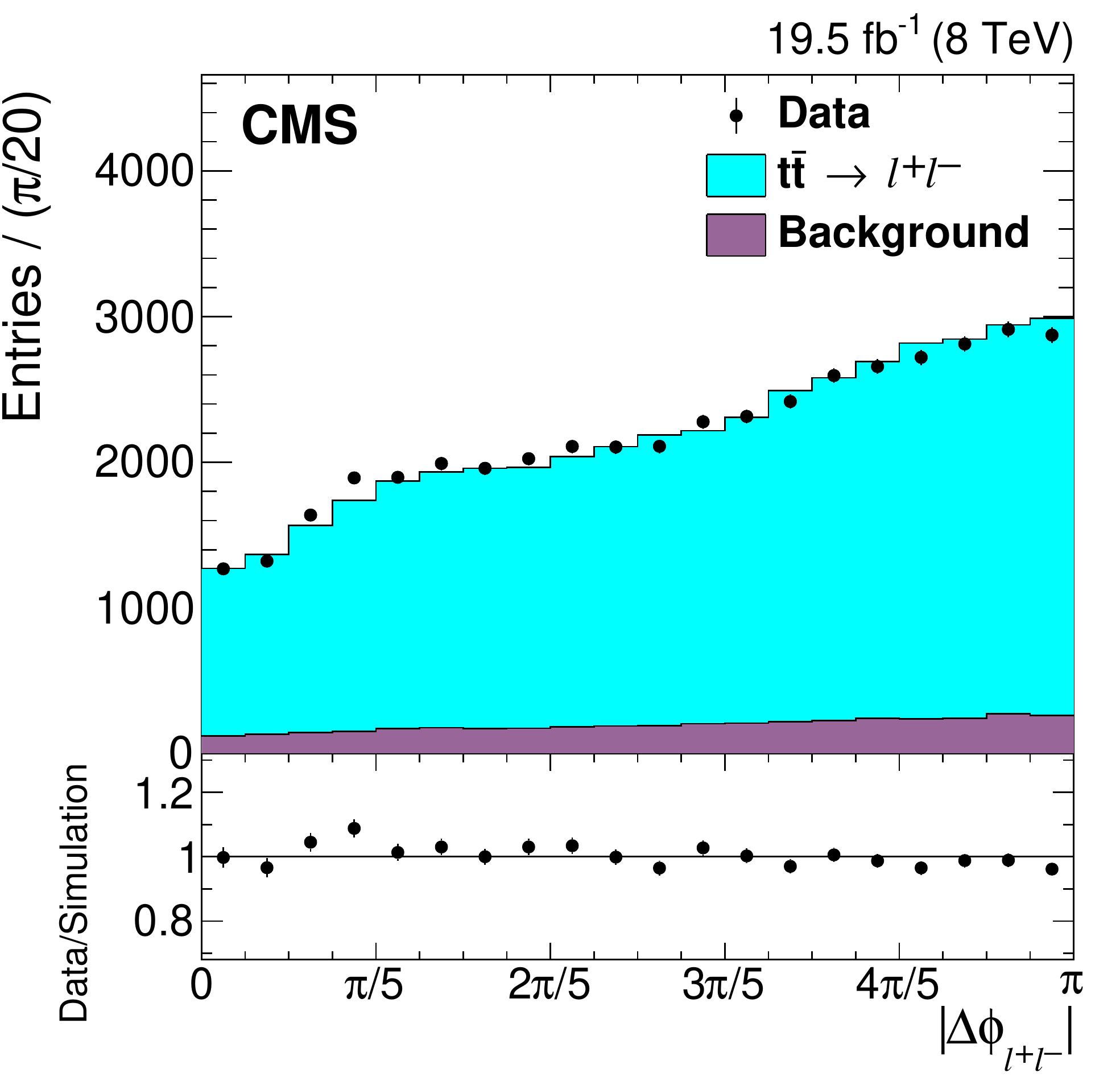}
\includegraphics[width=0.325\linewidth]{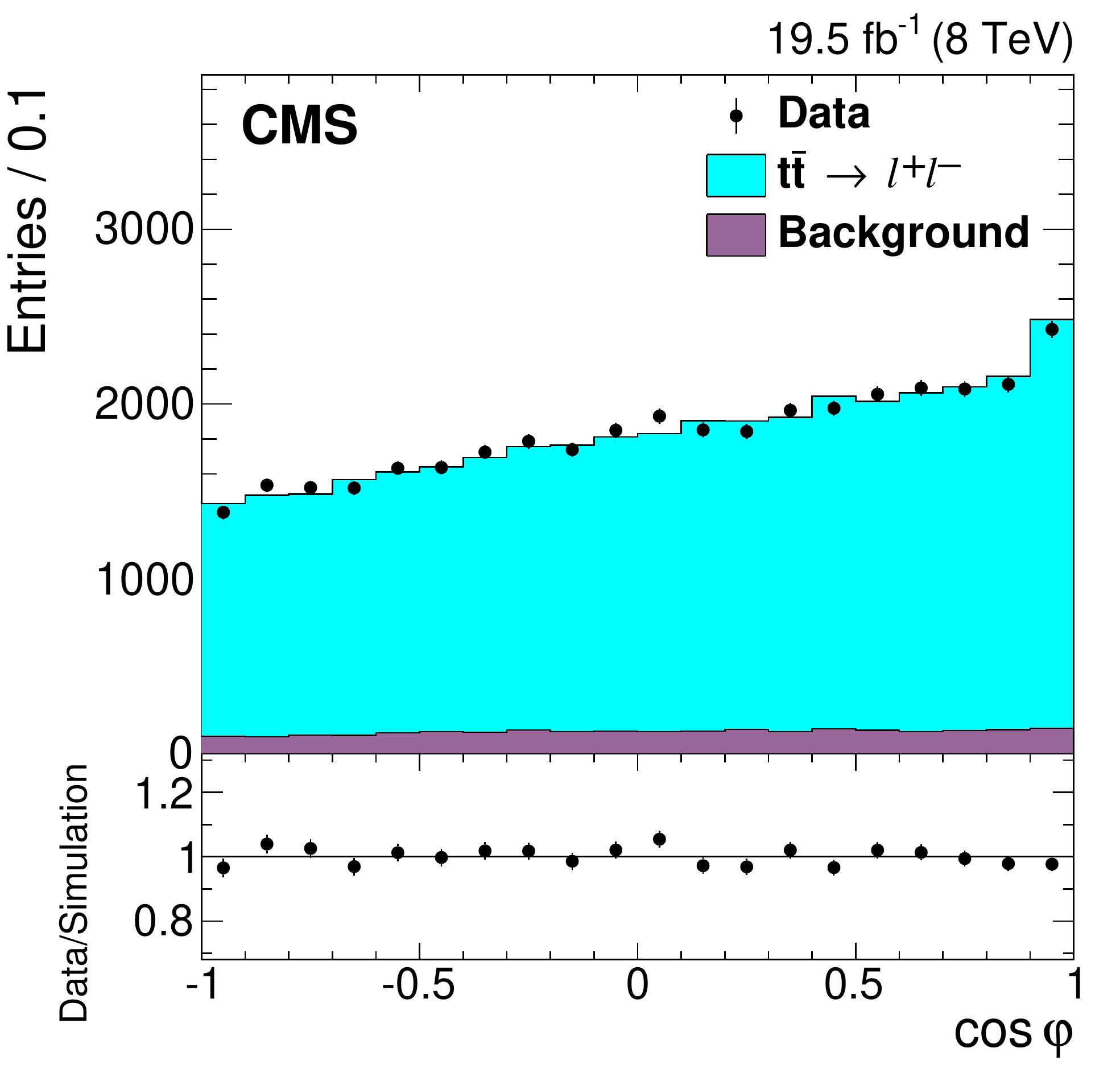}
\includegraphics[width=0.325\linewidth]{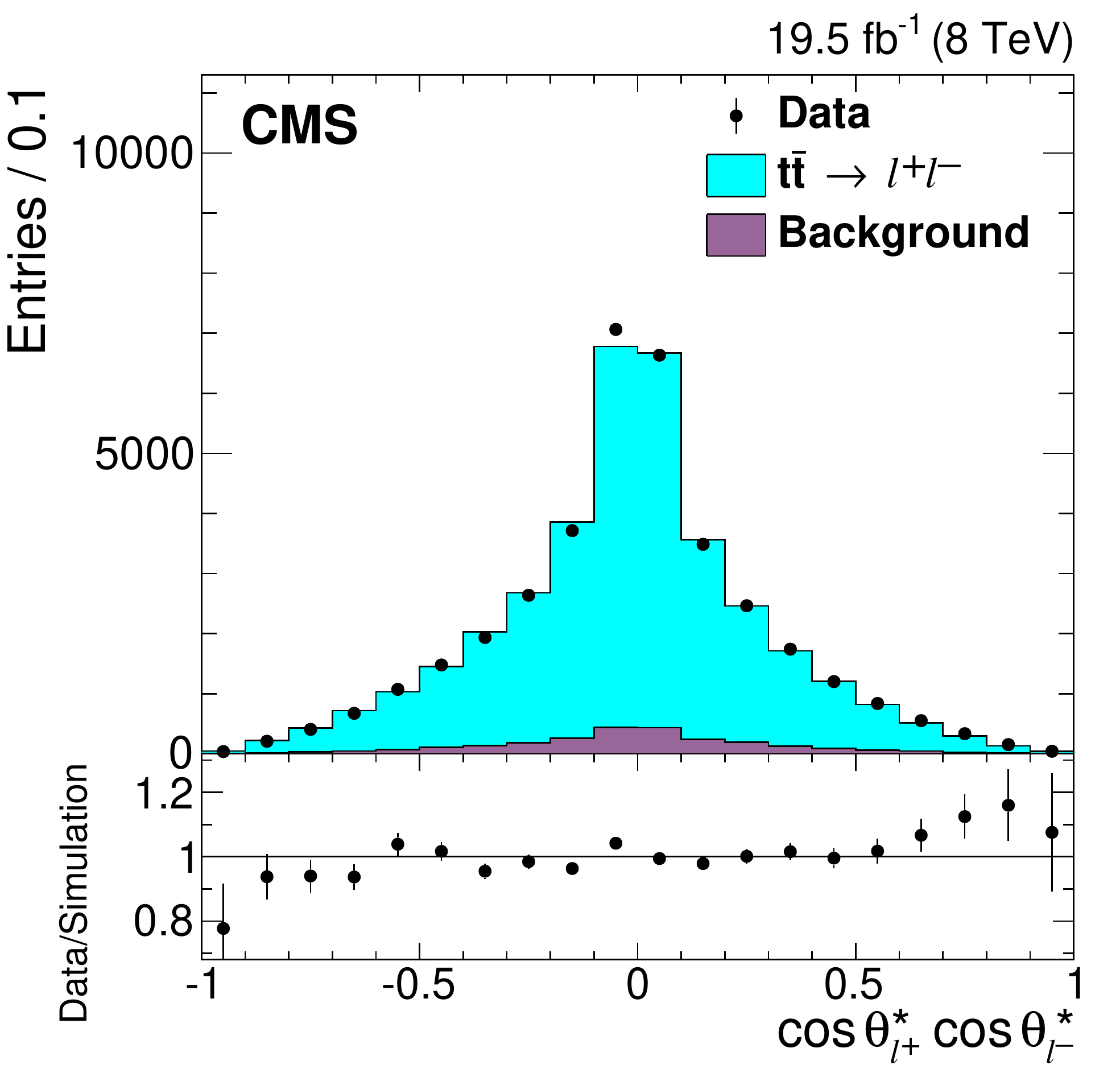}\\
\includegraphics[width=0.325\linewidth]{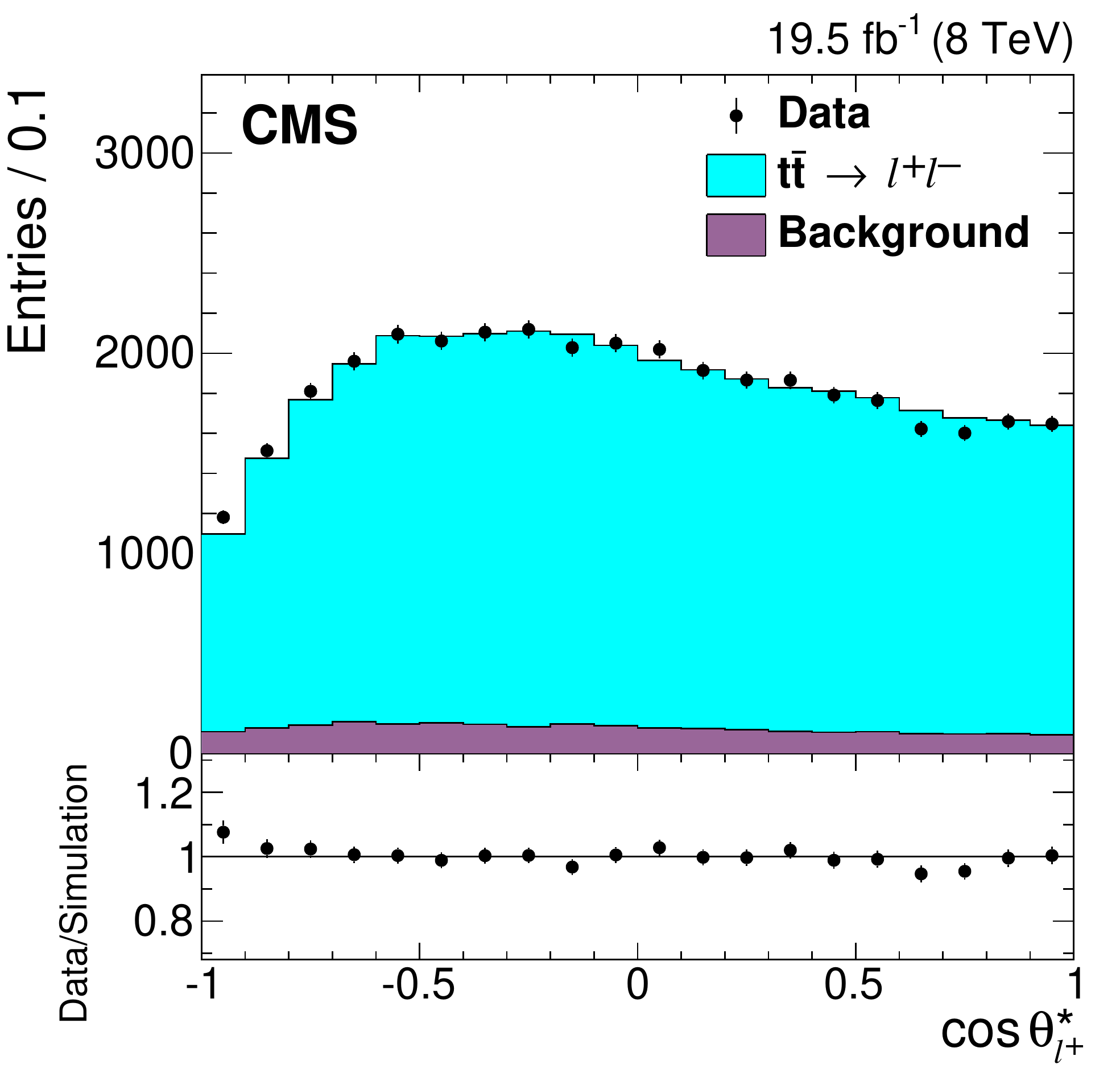}
\includegraphics[width=0.325\linewidth]{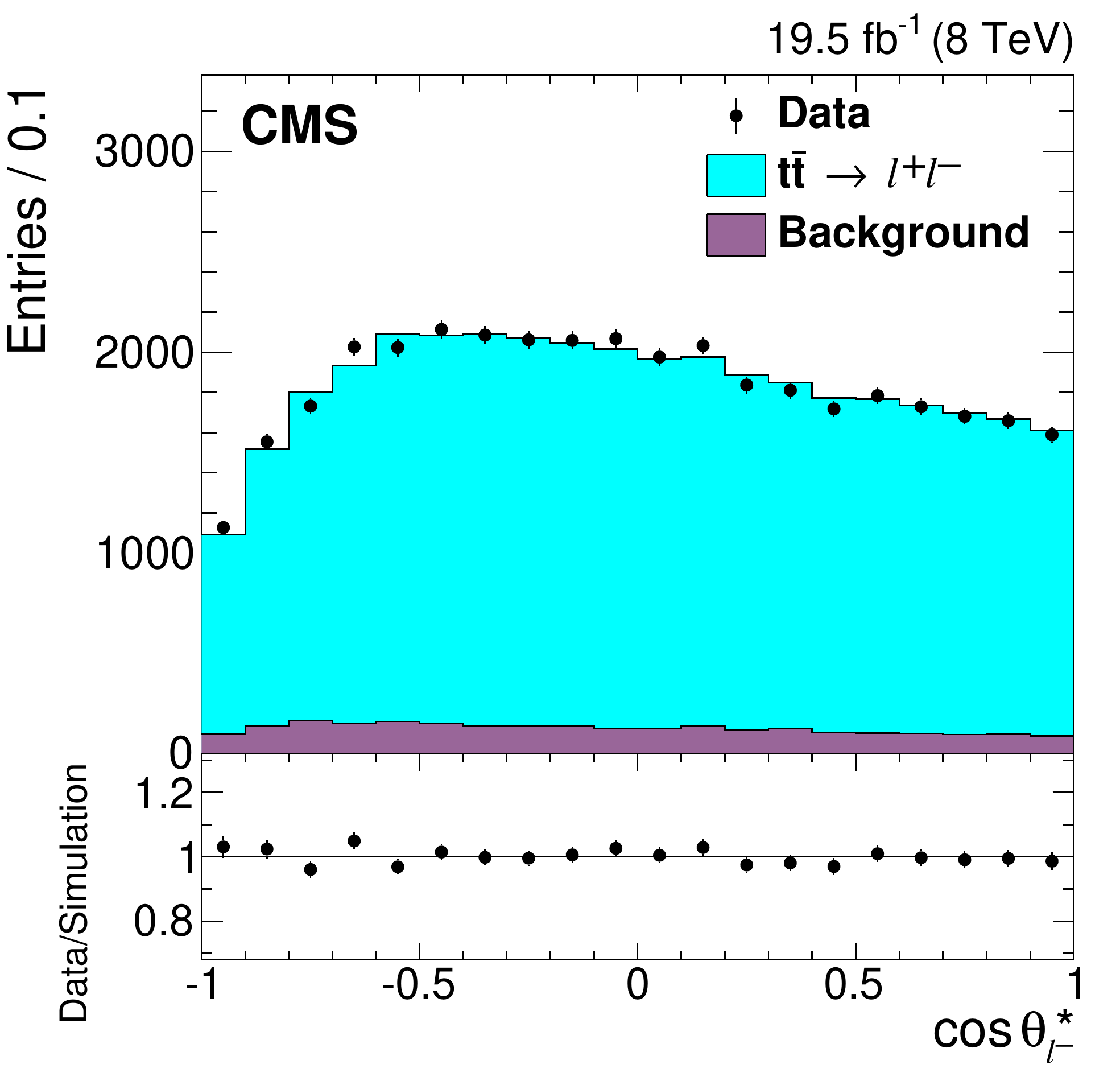}
\caption{\label{fig:preselcompasym}\protect
Reconstructed angular distributions from data (points) and simulation (histogram), with the expected signal (\ttll) and background distributions shown separately.
All three dilepton flavor combinations are included. The simulated signal yield is normalized to that of the background-subtracted data.
The vertical bars on the data points represent the statistical uncertainties.
The lower panels show the ratio of the numbers of events from data and simulation.
}
\end{figure*}

A comparison of the distributions for the reconstructed \ttbar\ system variables $M_{\ttbar}$, $y_{\ttbar}$, and $\ptttbar$ between data and simulation
 is shown in \fig~\ref{fig:preselcomp},
where the signal yield from the simulation has been normalized to the number of signal events in the data after background subtraction.
In general, the shapes of the distributions from data and simulation show reasonable agreement, with the small discrepancies covered by the systematic variations in the top quark \pt\ modeling, PDFs, and $\mu_\mathrm{R}$ and $\mu_\mathrm{F}$ values, which will be discussed in \secn~\ref{sec:systematics}.
A similar comparison of the angular distributions is shown in \fig~\ref{fig:preselcompasym}.
The corresponding inclusive asymmetry values, uncorrected for background, from the data and simulation are given in \tab~\ref{table:pres_asymm1}.

\begin{table}[!htpb]
\begin{center}
\topcaption{\label{table:pres_asymm1}
Values of the uncorrected inclusive asymmetry variables from simulation and data, prior to background subtraction.
The uncertainties shown are statistical.}
\begin{scotch}{l | X{6.6} X{6.6} }
  Reconstructed asymmetry
& \multicolumn{1}{c}{Simulation}
& \multicolumn{1}{c}{Data} \\
\hline
$A_{\Delta\phi}$  & 0.188 , 0.002 & 0.170 , 0.005  \\
$A_{\cos\varphi}$ & 0.114 , 0.003  & 0.109 , 0.005 \\
$A_{c_1 c_2}$ &  -0.050 , 0.003 & -0.049 , 0.005 \\
$A_{P+}$ & -0.026 , 0.003 & -0.032 , 0.005 \\
$A_{P-}$ & -0.022 , 0.003 & -0.028 , 0.005 \\
\end{scotch}
\end{center}
\end{table}

\section{Unfolding the distributions}
\label{sec:unfolding}
The observed angular distributions are distorted compared to the underlying distributions at the parton level (for which theoretical predictions exist)
by the detector acceptance and resolution and the trigger and event selection efficiencies.
To correct the data for these effects, we apply an unfolding procedure that
 yields the corrected $\abs{\Delta \phi_{\ell^+\ell^-}}$, $\cos\varphi$, $c_{1} c_{2}$, and $\cos\theta^{\star}_{\ell^\pm}$ distributions at the parton level.
In the context of theoretical calculations and parton-shower event generators, the parton-level top quark is defined before it decays, and its kinematics include the effects of recoil
from initial- and final-state radiation in the rest of the event and from final-state radiation from the top quark itself. The parton-level charged lepton, produced from the
decay of the intermediate \PW~boson, is defined before the lepton radiates any photons or the muon or tau lepton decays.

In order to unfold the observed distributions it is necessary to choose a binning scheme.
Aiming to have bins with widths
well matched to the reconstruction resolution and with approximately uniform event contents,
we select six bins for each parton-level angular distribution except that of $\Delta \phi_{\ell^+\ell^-}$.
This variable depends only on the lepton momentum measurements, not on the reconstruction of the \ttbar\ system,
and the superior resolution allows us to use 12 bins. For the
reconstruction-level distributions we use twice as many bins as for the parton-level distributions.

The background-subtracted distribution for each variable, considered as a vector $\vec{y}$, is related to the underlying
parton-level distribution $\vec{x}$ through the equation $\vec{y}=SA\vec{x}$,
where $A$ is a diagonal matrix describing the fraction (acceptance times efficiency) of all produced signal events that are expected to be selected in each of the measured bins, and
$S$ is a nondiagonal ``smearing'' matrix describing the migration of events between bins caused by imperfect detector resolution
and reconstruction techniques. The $A$ and $S$ matrices are constructed using simulated \MCATNLO\ \ttbar\ events.
The smearing in $\cos\varphi$, $c_{1} c_{2}$, and $\cos\theta^{\star}_{\ell^\pm}$ can be large in some events because of the uncertainties in the reconstruction of the \ttbar\ kinematic quantities, but the smearing matrices are
still predominantly diagonal.
The smearing matrix for $\abs{\Delta \phi_{\ell^+\ell^-}}$
is nearly diagonal because of the excellent angular resolution of the lepton momentum measurements.

To determine the parton-level angular distribution in data,
we employ a regularized unfolding algorithm implemented in the {\sc TUnfold} package~\cite{1748-0221-7-10-T10003}.
 The effects of large statistical fluctuations in the algorithm are
greatly reduced by introducing a term in the unfolding procedure that regularizes the output distribution
based on the curvature of the simulated signal distribution.
In general, unfolding introduces negative correlations between adjacent bins, while regularization introduces positive correlations, and
the regularization strength is optimized by minimizing the average global correlation coefficient in the unfolded distribution.
The regularization strength obtained here is relatively weak, contributing at the $10\%$ level to the total $\chi^2$ minimized by the algorithm.

After unfolding, each distribution is normalized to unit area to give the normalized differential cross section for each variable.
We use an analogous unfolding procedure to measure the normalized double-differential cross section, using three bins of $M_{\ttbar}$, $\abs{y_{\ttbar}}$, and $\ptttbar$ for each variable.
The full covariance matrix is used in the evaluation of the statistical uncertainty in the asymmetry measured from each distribution.

\section{Systematic uncertainties}	
\label{sec:systematics}

The systematic uncertainties coming from the detector performance and the modeling of the signal and background processes
are evaluated from the
difference between the nominal measurement and that obtained by repeating the unfolding procedure using simulated events with the appropriate systematic variation.

The uncertainty from the jet energy scale (JES) corrections affects the \ttbar\ final-state reconstruction, as well as the event selection.
It is estimated by varying the energies of jets within their uncertainties~\cite{Chatrchyan:2011ds}, and propagating this to the \MET\ value.
Similarly, the jet energy resolution is varied by 2--5\%, depending on the
$\eta$ of the jet~\cite{Chatrchyan:2011ds}, and
the electron energy scale is varied
by ${\pm}0.6\%$ (${\pm}1.5\%$) for barrel (endcap) electrons (the uncertainty in muon energies is negligible), as estimated from comparisons between measured and simulated Drell--Yan events~\cite{1748-0221-8-09-P09009}.

The uncertainty in the background contribution is obtained by
varying the normalization of each background component by the uncertainties described in \secn~\ref{Sec:BkgEst}.

Many of the signal modeling and simulation uncertainties are evaluated by using weights to vary the \MCATNLO \ttbar\ sample:
the simulated pileup multiplicity distribution is changed within its uncertainty;
the correction factors between data and simulation for the $\PQb$~tagging~\cite{ref:btag}, trigger, and lepton selection efficiencies are shifted up and down by their uncertainties;
and the PDFs are varied using the PDF4LHC procedure~\cite{Alekhin:2011sk,pdf4lhcInterim}.
Previous CMS studies~\cite{toppT,Khachatryan:2015oqa} have shown that the \pt\ distribution of the top quark measured from data is softer than that in the NLO simulation of \ttbar\ production.
Since the origin of the discrepancy is not fully understood, the change in the measurement when reweighting
the \MCATNLO \ttbar\ sample to match the top quark \pt\ spectrum in data is taken as a systematic uncertainty associated with signal modeling.

The remaining signal modeling uncertainties are separately evaluated with dedicated \ttbar\ samples:
$\mu_\mathrm{R}$ and $\mu_\mathrm{F}$ are varied together up and down by a factor of 2;
the top quark mass is varied by ${\pm}1\GeV$, to be consistent with the
uncertainty used in other CMS measurements with the $\sqrt{s}=8\TeV$ data set (the
effect on the total systematic uncertainty of using the reduced uncertainty from the recent CMS combined $m_{\cPqt}$
measurement~\cite{CMSNewMass} would be negligible);
and
the $S$ matrix is rederived from a \ttbar\ sample generated with \POWHEG and \PYTHIA, while the $A$ matrix is unchanged, in order to
estimate the difference in hadronization modeling between \HERWIG and \PYTHIA.
To avoid underestimation of systematic uncertainties caused by statistical fluctuations, which can be significant in the estimates evaluated using dedicated \ttbar\ samples, for each source of uncertainty the maximum of the estimated systematic uncertainty and the statistical uncertainty in that estimate is taken as the final systematic uncertainty.

The uncertainty in the unfolding procedure is dominated by the statistical uncertainty arising from the finite number of events in the \MCATNLO \ttbar\ sample.
The uncertainty owing to the unfolding regularization is evaluated by using the
reconstucted distribution of a variable in data to reweight the
corresponding simulated signal distribution used to regularize the
curvature of the unfolded distribution. Using this method, the
regularization uncertainty is found to be negligible for all measurements.

\begin{table*}[!htpb]															
\begin{center}															
\topcaption{\label{tab:asyms_sys} Sources and values of the systematic uncertainties in the inclusive asymmetry variables.}
\begin{scotch}{ l | r r r r r }															
Asymmetry variable		&	\multicolumn{1}{c}{$A_{\Delta\phi}$}	&	\multicolumn{1}{c}{$A_{\cos\varphi}$}  &	 \multicolumn{1}{c}{$A_{c_1 c_2}$}	
	&	\multicolumn{1}{c}{$A_{P}$}  &	\multicolumn{1}{c}{$A_{P}^{\mathrm{CPV}}$}     \\	[0.1ex]
\hline															
\hline															
\multicolumn{6}{c}{Experimental systematic uncertainties}	\\														
\hline															
Jet energy scale                  & $0.001$   & $0.005$   & $0.007$   & $0.018$   & $0.001$   \\
Jet energy resolution             & ${<}0.001$   & $0.001$   & $0.002$   & $0.003$   & $0.002$   \\
Lepton energy scale               & $0.001$   & $0.002$   & $0.005$   & $0.003$   & ${<}0.001$   \\
Background                        & $0.001$   & $0.001$   & $0.001$   & $0.002$   & ${<}0.001$   \\
Pileup                            & ${<}0.001$   & ${<}0.001$   & ${<}0.001$   & ${<}0.001$   & ${<}0.001$   \\
$\PQb$~tagging efficiency              & ${<}0.001$   & $0.001$   & $0.001$   & $0.001$   & $0.001$   \\
Lepton selection                  & $0.001$   & ${<}0.001$   & ${<}0.001$   & $0.002$   & ${<}0.001$   \\
\hline
\hline
\multicolumn{6}{c}{\ttbar\ modeling uncertainties}	\\														
\hline															
Parton distribution functions     & $0.004$   & $0.005$   & $0.005$   & $0.001$   & ${<}0.001$   \\
Top quark \pt\                    & $0.011$   & $0.006$   & $0.006$   & $0.004$   & ${<}0.001$   \\
Fact. and renorm. scales          & $0.002$   & $0.003$   & $0.005$   & $0.002$   & $0.002$   \\
Top quark mass                    & $0.001$   & $0.001$   & $0.007$   & $0.008$   & $0.001$   \\
Hadronization                     & $0.001$   & $0.004$   & $0.005$   & $0.019$   & $0.003$   \\
\hline															
\hline															
Unfolding (simulation statistical) & $0.002$   & $0.005$   & $0.006$   & $0.003$   & $0.003$   \\
Unfolding (regularization)        & ${<}0.001$   & ${<}0.001$   & ${<}0.001$   & ${<}0.001$   & ${<}0.001$   \\
\hline															
\hline															
Total systematic uncertainty      &  $0.012$ &  $0.012$ &  $0.016$ &  $0.028$ &  $0.005$ \\
\end{scotch}																													
\end{center}															
\end{table*}				

The systematic uncertainties in the inclusive asymmetry variables obtained from the unfolded distributions are summarized in \tab~\ref{tab:asyms_sys}.
The systematic uncertainties are evaluated for each bin of the unfolded distributions,
from which the covariance matrix is constructed, assuming 100\% correlation or anticorrelation between bins for each individual source of uncertainty.
The total systematic uncertainty is calculated by adding in quadrature the listed uncertainties.

For $A_{\Delta\phi}$, the top quark \pt modeling uncertainty dominates; this arises from the dependence of the $\abs{\Delta \phi_{\ell^+\ell^-}}$ distribution shape on the top quark \pt (through the spin correlations and event kinematics); that, in turn, introduces
a significant dependence of the acceptance correction on the top quark \pt.
For $A_{P}$, the JES and hadronization systematic uncertainties are dominant. Both affect the reconstructed $\PQb$~quark jet energy, and can therefore bias the boost from the laboratory frame to the top quark center-of-mass frame,
and thus the measurement of $\cos\theta^{\star}_{\ell^\pm}$. For similar reasons, the same two uncertainties are large for $A_{c_1 c_2}$ and $A_{\cos\varphi}$, which are also significantly affected by the top quark \pt modeling uncertainty
through its effect on the spin correlations. For $A_{P}^{\mathrm{CPV}}$, the similar systematic uncertainties in $A_{P+}$ and $A_{P-}$ largely cancel
when $A_{P-}$ is subtracted from $A_{P+}$;
the remaining contributions to the systematic uncertainty are dominated by the statistical uncertainty in the simulation.

\section{Results}

\subsection{Unfolded distributions}
\label{sec:results}

\begin{figure*}[!htpb]
\begin{center}

\includegraphics[width=0.48\linewidth]{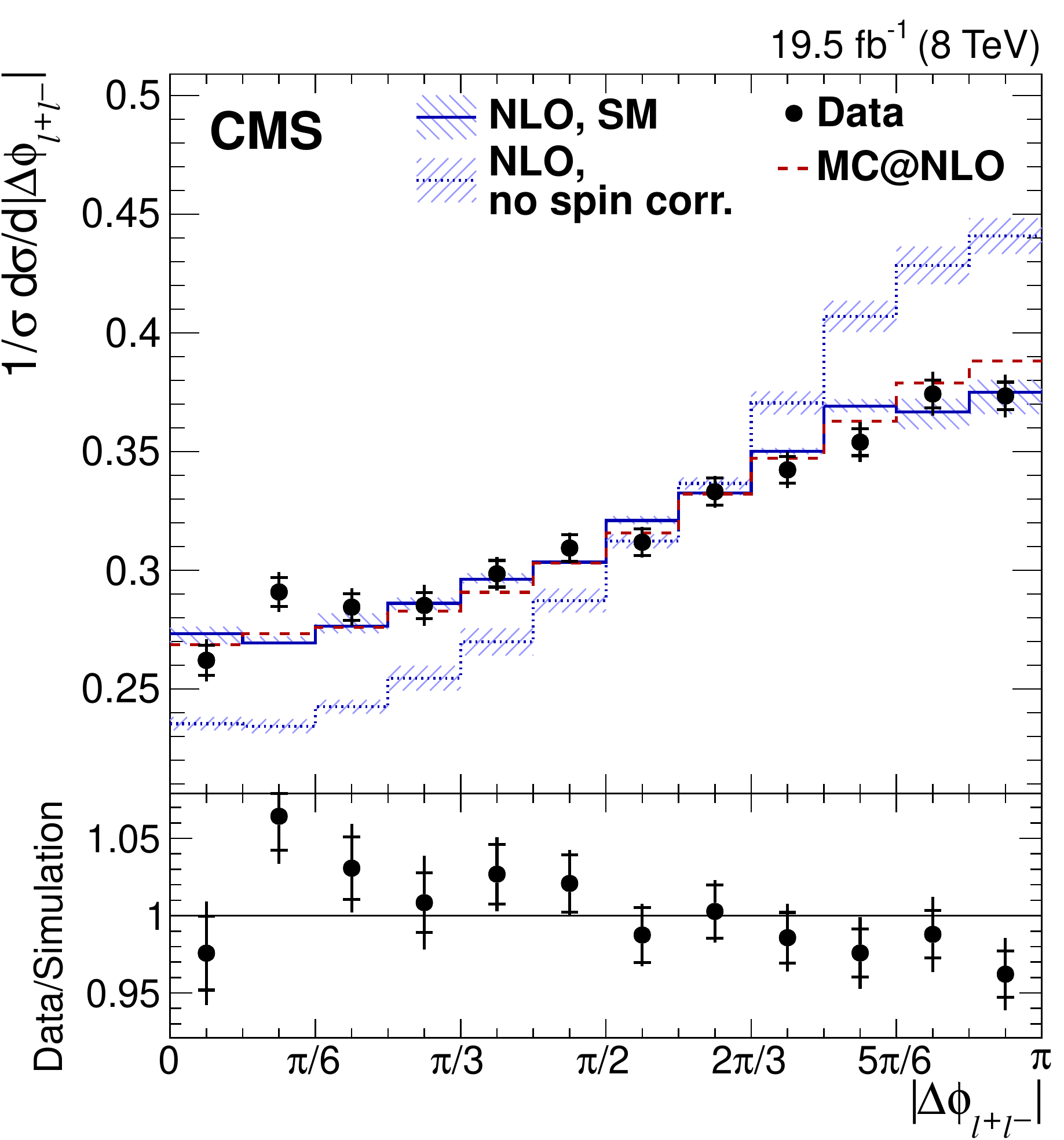}
\includegraphics[width=0.48\linewidth]{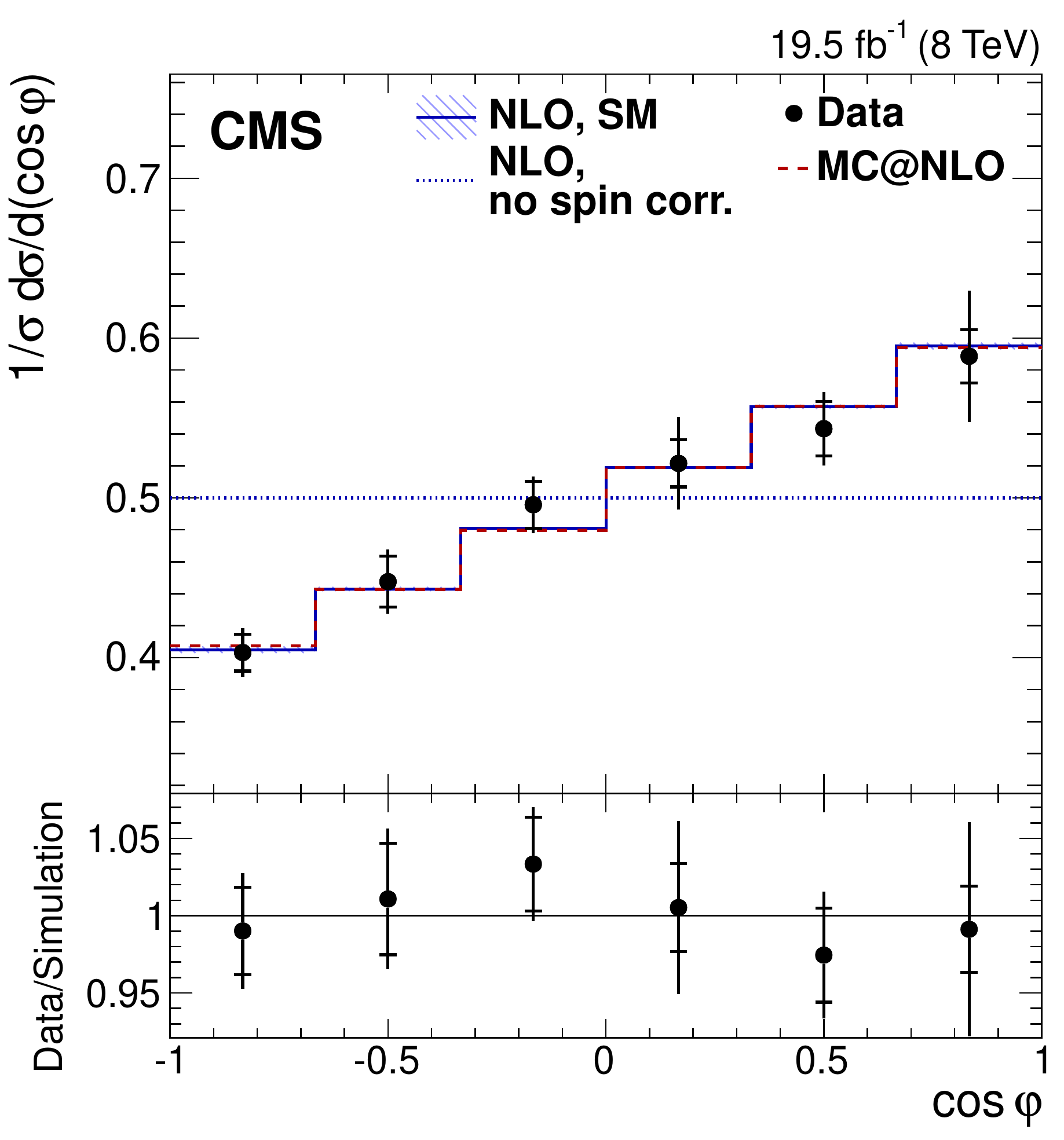}\\
\includegraphics[width=0.48\linewidth]{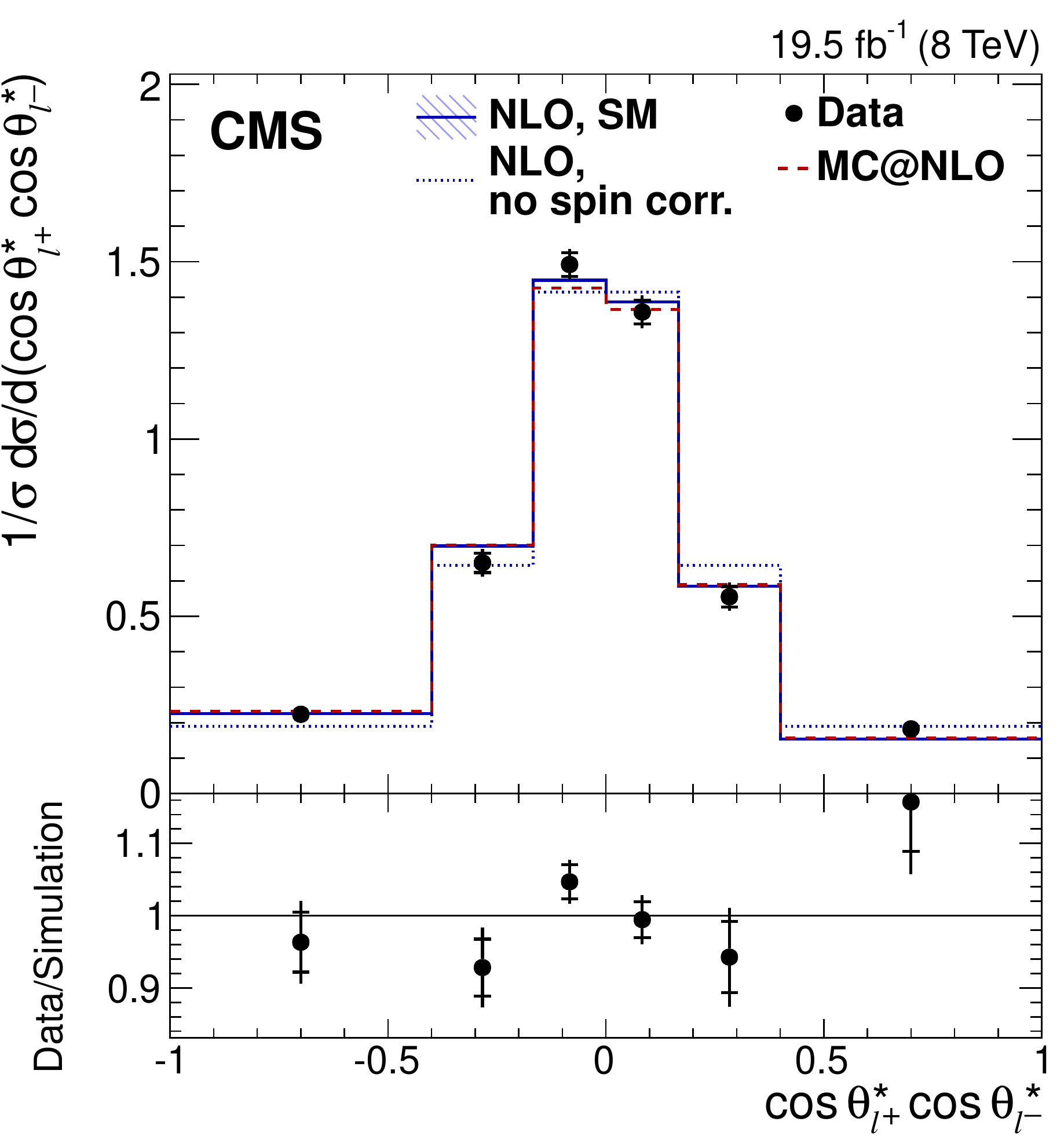}
\includegraphics[width=0.48\linewidth]{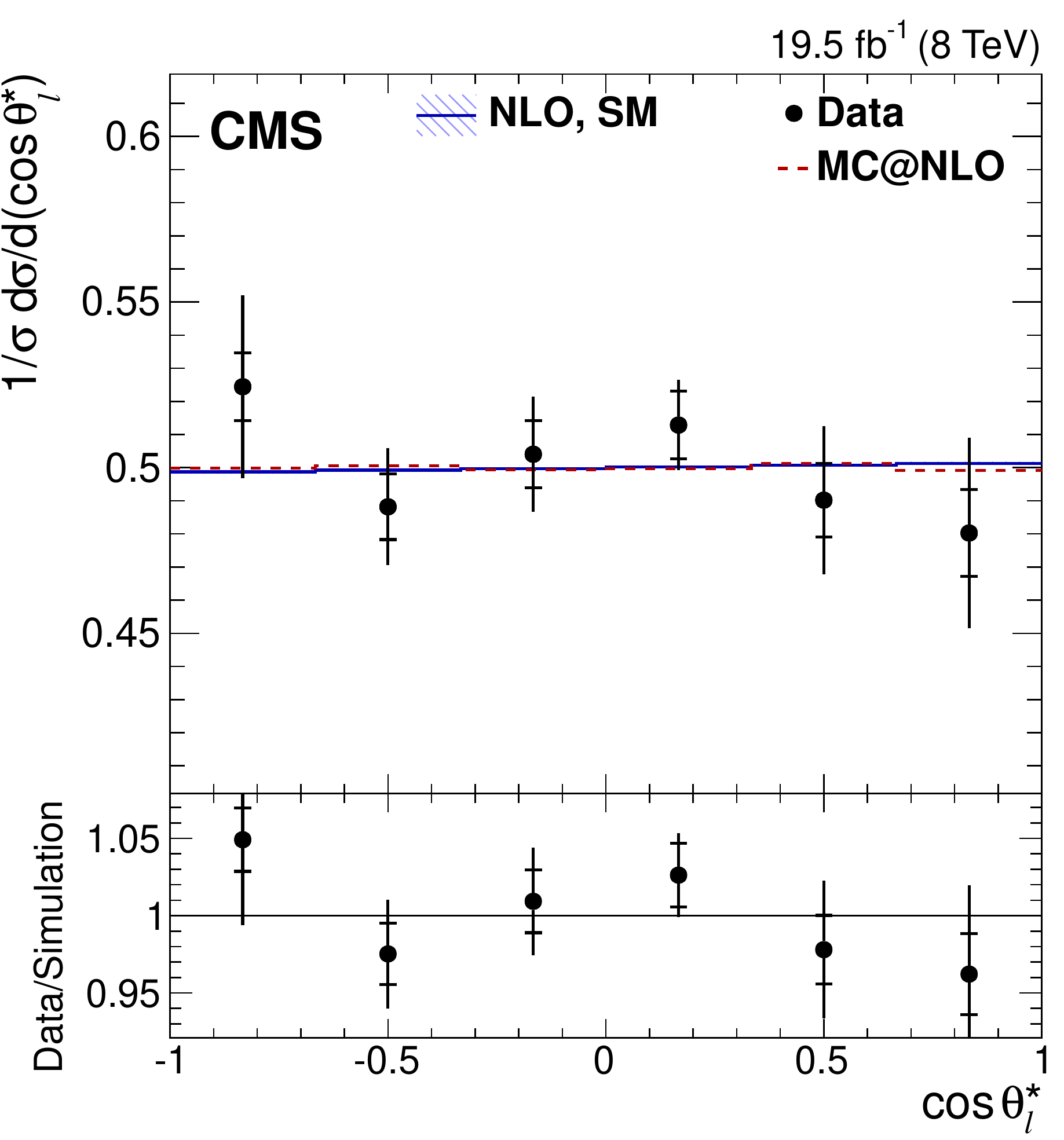}
\caption{\label{fig:ResultsUnfolded}\protect
Normalized differential cross section as a function of
$\abs{\Delta \phi_{\ell^+\ell^-}}$, $\cos\varphi$, $\cos\theta^{\star}_{\ell^+} \cos\theta^{\star}_{\ell^-}$, and $\cos\theta^{\star}_\ell$ from data (points); parton-level predictions from \MCATNLO\ (dashed histograms); and theoretical predictions at \NLO~\cite{theoretical,Bernreuther2013115} with (SM) and without (no spin corr.) spin correlations (solid and dotted histograms, respectively).
For the $\cos\theta^{\star}_\ell$ distribution, CP conservation is assumed in the combination of the $\cos\theta^{\star}_{\ell^\pm}$ measurements from positively and negatively charged leptons.
The ratio of the data to the \MCATNLO\ prediction is shown in the lower panels.
The inner and outer vertical bars on the data points represent the statistical and total uncertainties, respectively.
The hatched bands represent variations of $\mu_\mathrm{R}$ and $\mu_\mathrm{F}$ simultaneously up and down by a factor of 2.
}
\end{center}
\end{figure*}

The background-subtracted, unfolded, and normalized-to-unit-area angular distributions for the selected data events are shown in \fig~\ref{fig:ResultsUnfolded}, along with the parton-level predictions obtained with the \MCATNLO\ event generator
and from calculations at NLO in the strong and weak gauge couplings for \ttbar\ production, with and without spin correlations~\cite{theoretical,Bernreuther2013115}.

The measured asymmetries, obtained from the angular distributions unfolded to the parton level, are presented with their statistical and systematic uncertainties in \tab~\ref{tab:ResultsUnfolded}, where they are compared to predictions from \MCATNLO and the \NLO calculations.
Correlations between the contents of different bins, introduced by the unfolding process and from the systematic uncertainties, are accounted for in the calculation of the experimental uncertainties.
The uncertainties in the \NLO predictions come from varying $\mu_\mathrm{R}$ and $\mu_\mathrm{F}$ simultaneously up and down by a factor of 2.
For $A_{\cos\varphi}$ and $A_{c_1 c_2}$, these scale uncertainties are summed in quadrature with the difference between the \NLO predictions from \reference~\cite{Bernreuther2013115} when the ratio in the calculation is expanded in powers of the strong coupling constant
and when the numerator and denominator are evaluated separately.

\begin{table*}[!htpb]
\centering
\topcaption{\label{tab:ResultsUnfolded}
Inclusive asymmetry measurements obtained from the angular distributions unfolded to the parton level, and the parton-level predictions from the \MCATNLO\ simulation and from \NLO calculations with (SM) and without (no spin corr.) spin correlations~\cite{theoretical,Bernreuther2013115}.
For the data, the first uncertainty is statistical and the second is systematic. For the \MCATNLO\ results and \NLO calculations, the uncertainties are statistical and theoretical, respectively.
}
\begin{scotch}{l | X{6.12} X{6.5} X{6.5} X{5.5} }
\multicolumn{1}{c|}{Asymmetry} & \multicolumn{1}{c}{Data} & \multicolumn{1}{c}{\MCATNLO} & \multicolumn{1}{c}{\NLO,} & \multicolumn{1}{c}{\NLO,} \\
\multicolumn{1}{c|}{variable} & \multicolumn{1}{c}{(unfolded)} & \multicolumn{1}{c}{simulation} & \multicolumn{1}{c}{SM} & \multicolumn{1}{c}{no spin corr.} \\
\hline & & & & \\ [-2.4ex]
$~~~~~A_{\Delta\phi}$	&  0.094 ,   0.005 \pm 0.012  &   0.113 , 0.001   &  \multicolumn{1}{c}{$0.107\:^{+\:0.006}_{-\:0.009}$} &  \multicolumn{1}{c}{$0.202\:^{+\:0.006}_{-\:0.009}$}\\
$~~~~~A_{\cos\varphi}$	&  0.102 , 0.010 \pm 0.012  &   0.114 , 0.001   &  0.114 , 0.006 & \multicolumn{1}{c}{~~0} \\
$~~~~~A_{c_1 c_2}$			&  -0.069 ,  0.013 \pm 0.016  &   -0.081, 0.001   &  -0.080, 0.004 & \multicolumn{1}{c}{~~0} \\
$~~~~~A_{P}$			&  -0.011 ,  0.007 \pm 0.028  & \multicolumn{1}{c}{~~0}  &  0.002 , 0.001 & \multicolumn{1}{c}{~~\NA} \\
$~~~~~A_{P}^{\mathrm{CPV}}$	&  0.000 ,  0.006 \pm 0.005  &  \multicolumn{1}{c}{~~0}  & \multicolumn{1}{c}{~~0} & \multicolumn{1}{c}{~~\NA} \\ [0.1ex]
\end{scotch}
\end{table*}

Using the relationships between the asymmetry variables and spin correlation coefficients given in \secn~\ref{sec:intro}, we find $C_{\mathrm{hel}} = 0.278 \pm 0.084$ and $D = 0.205 \pm 0.031$, where the uncertainties include the statistical and systematic components added in quadrature. Similarly, the CP-conserving and CP-violating components of the top quark polarization are found to be $P = -0.022 \pm 0.058$ and $P^{\mathrm{CPV}} = 0.000 \pm 0.016$, respectively.
All measurements are consistent with the expectations of the SM.

The \NLO predictions for $\abs{\Delta \phi_{\ell^+\ell^-}}$, $\cos\varphi$, and $c_{1} c_{2}$ with and without spin correlations in \tab~\ref{tab:ResultsUnfolded} are used to translate the measurements into determinations of $f_{\mathrm{SM}}$, the strength of the spin correlations relative to the SM prediction, with $f_{\mathrm{SM}}=1$ corresponding to the SM and $f_{\mathrm{SM}}=0$ corresponding to uncorrelated events. The measurements of $f_{\mathrm{SM}}$ are shown in \tab~\ref{tab:fSM} and are derived under the assumption that the $A$ matrix used for the unfolding is independent of spin correlations. This is found to give conservative estimates for the experimental uncertainties.

\begin{table*}[!htpb]
\centering
\topcaption{\label{tab:fSM}
Values of $f_{\mathrm{SM}}$, the strength of the measured spin correlations relative to the SM prediction, derived from the numbers in \tab~\ref{tab:ResultsUnfolded}.
The last row shows an additional measurement of $f_{\mathrm{SM}}$ made from the projection in $\abs{\Delta \phi_{\ell^+\ell^-}}$ of the normalized double-differential cross section as a function of $\abs{\Delta \phi_{\ell^+\ell^-}}$ and $M_{\ttbar}$.
The uncertainties shown are statistical, systematic, and theoretical, respectively. The total uncertainty in each result, found by adding the individual uncertainties in quadrature, is shown in the last column.
}
\begin{scotch}{l | X{4.12} X{-1}  }
  Variable
& \multicolumn{1}{c}{$f_{\mathrm{SM}} \pm \text{(stat)} \pm \text{(syst)} \pm \text{(theor)}$}
& \multicolumn{1}{c}{Total uncertainty} \\
\hline & & \\ [-2.4ex]
$A_{\Delta\phi}$						& 1.14 , 0.06 \pm 0.13 \:^{+\:0.08}_{-\:0.11} & \multicolumn{1}{c}{$^{+\:0.16}_{-\:0.18}$} \\ [0.3ex]
$A_{\cos\varphi}$						& 0.90 , 0.09 \pm 0.10 \pm 0.05 &  \multicolumn{1}{c}{$\,\pm\;\!0.15$} \\ [0.3ex]
$A_{c_1 c_2}$								& 0.87 , 0.17 \pm 0.21 \pm 0.04 &  \multicolumn{1}{c}{$\,\pm\;\!0.27$} \\ [0.3ex]
$A_{\Delta\phi}$ (vs. $M_{\ttbar}$) 	& 1.12 , 0.06 \pm 0.08 \:^{+\:0.08}_{-\:0.11} & \multicolumn{1}{c}{$^{+\:0.12}_{-\:0.15}$} \\ [0.3ex]
\end{scotch}
\end{table*}

\begin{figure*}[!htpb]
\centering
\includegraphics[width=0.300\linewidth]{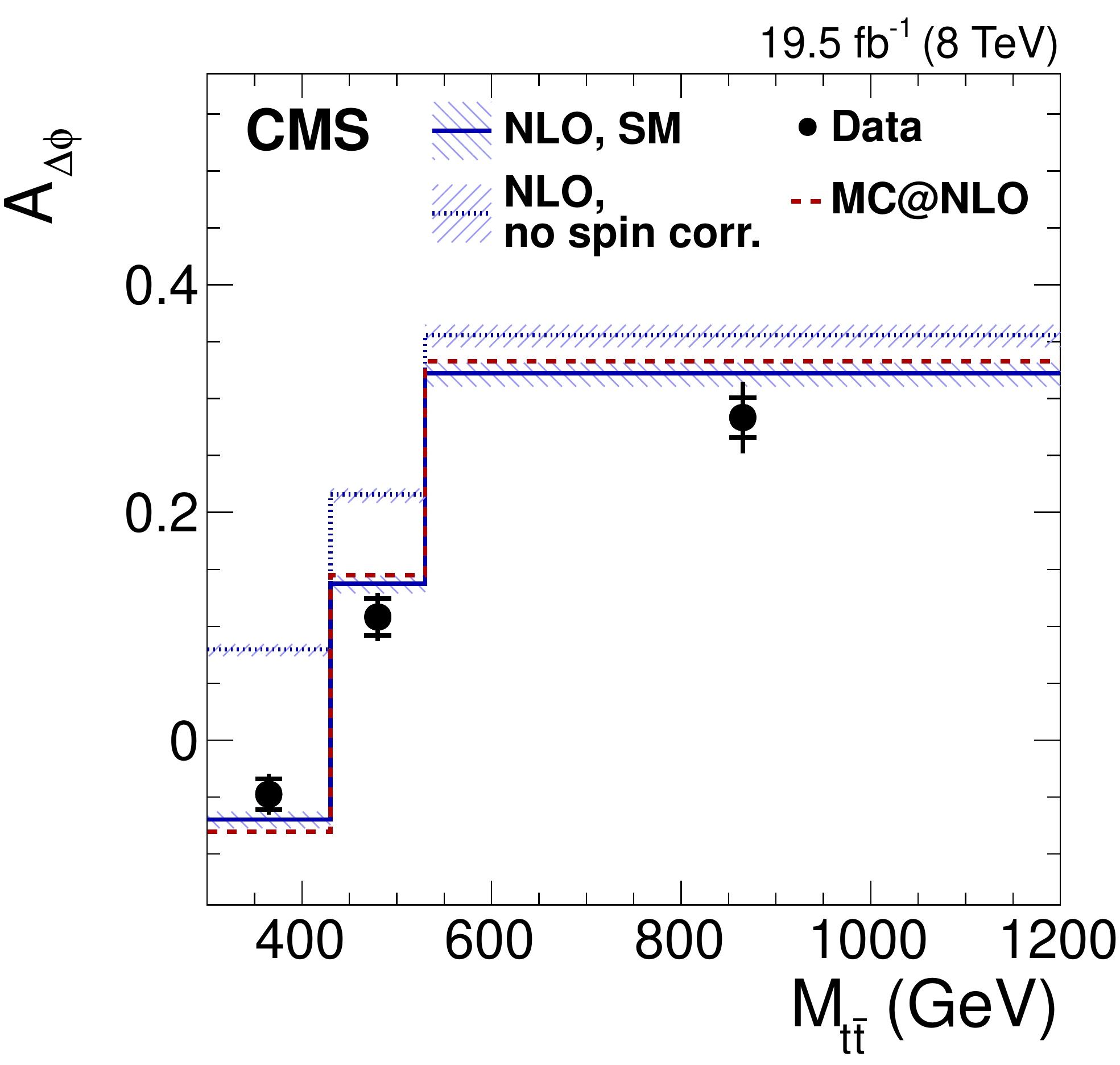}
\includegraphics[width=0.300\linewidth]{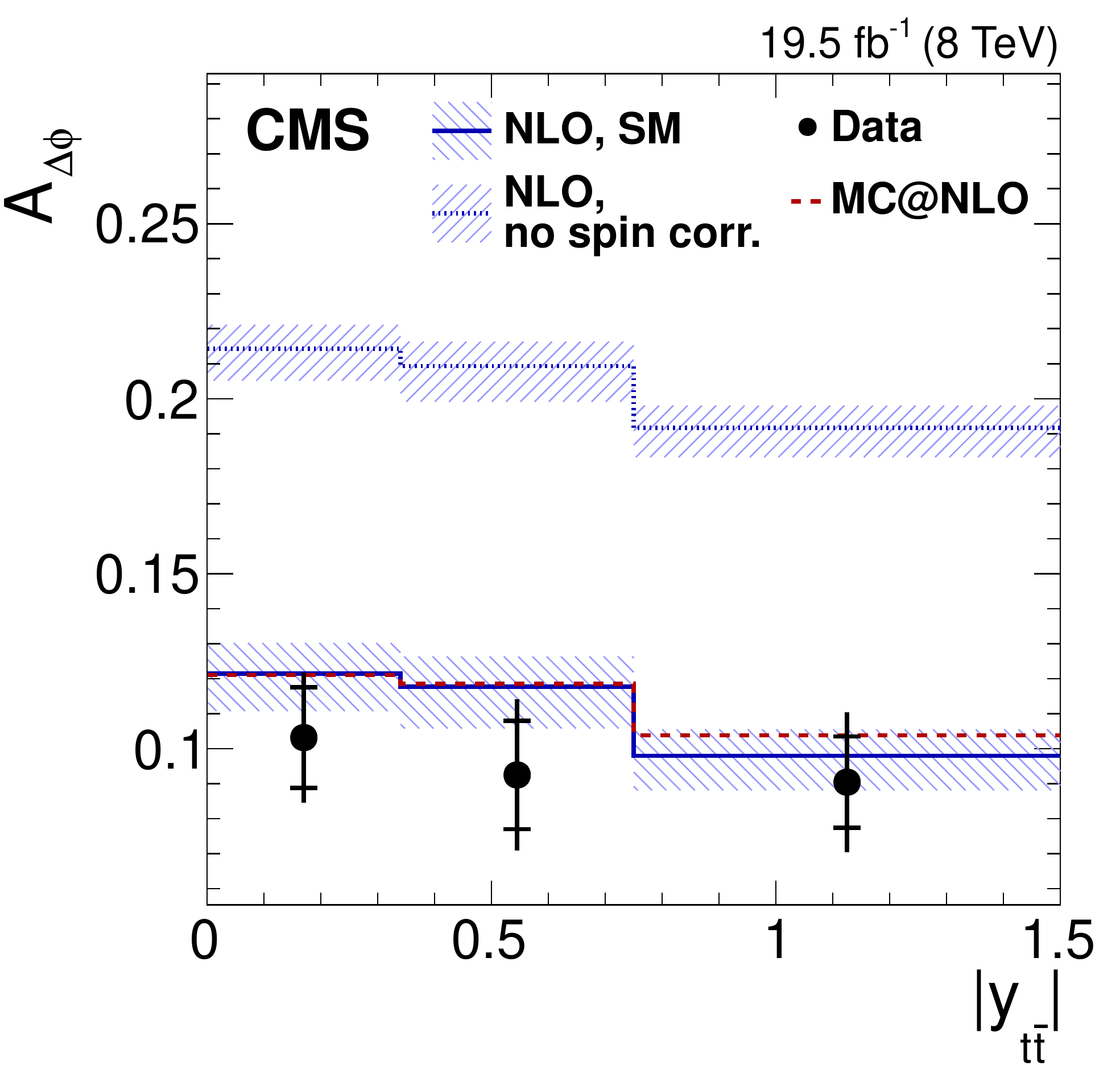}
\includegraphics[width=0.300\linewidth]{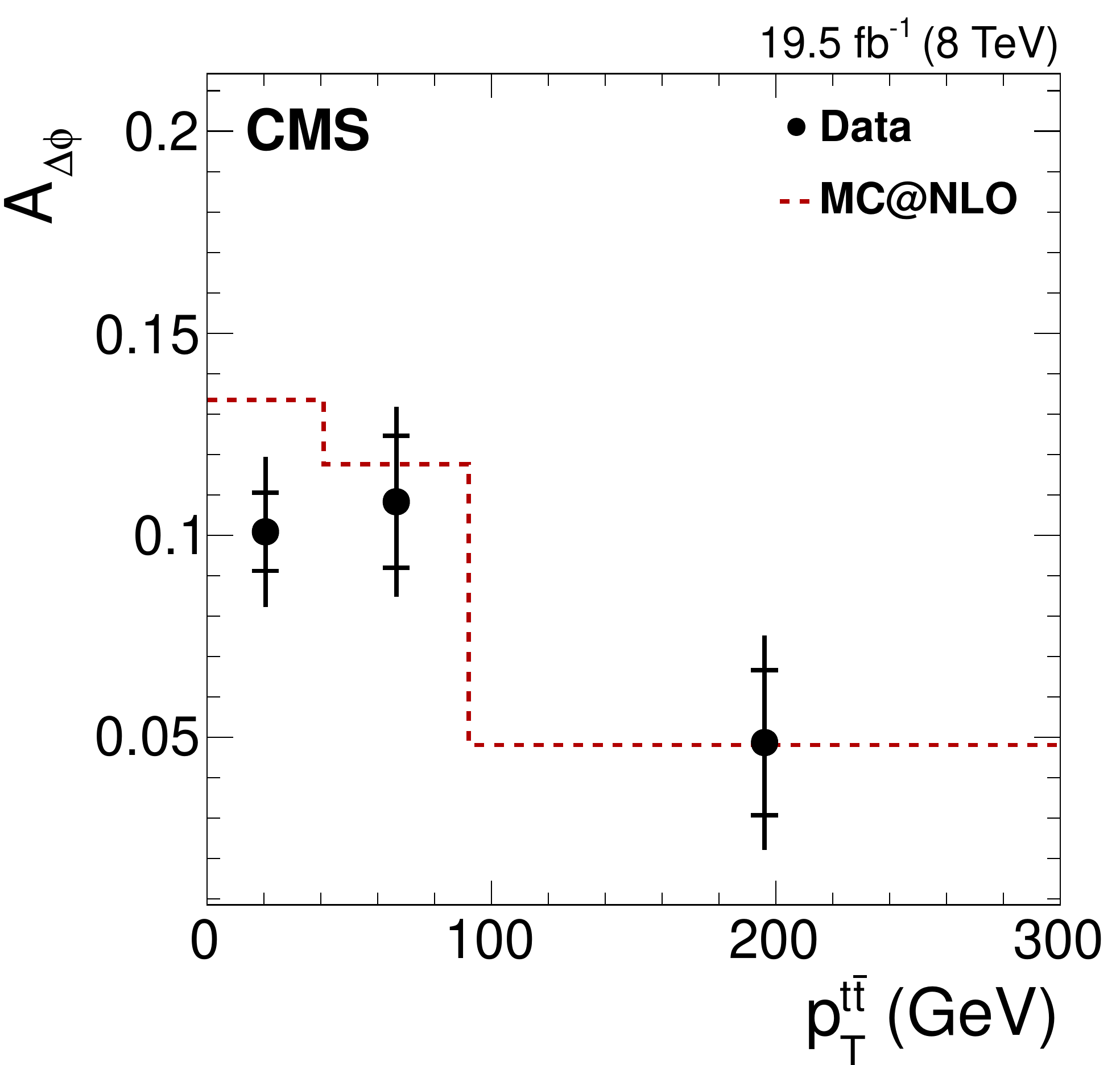}\\
\includegraphics[width=0.300\linewidth]{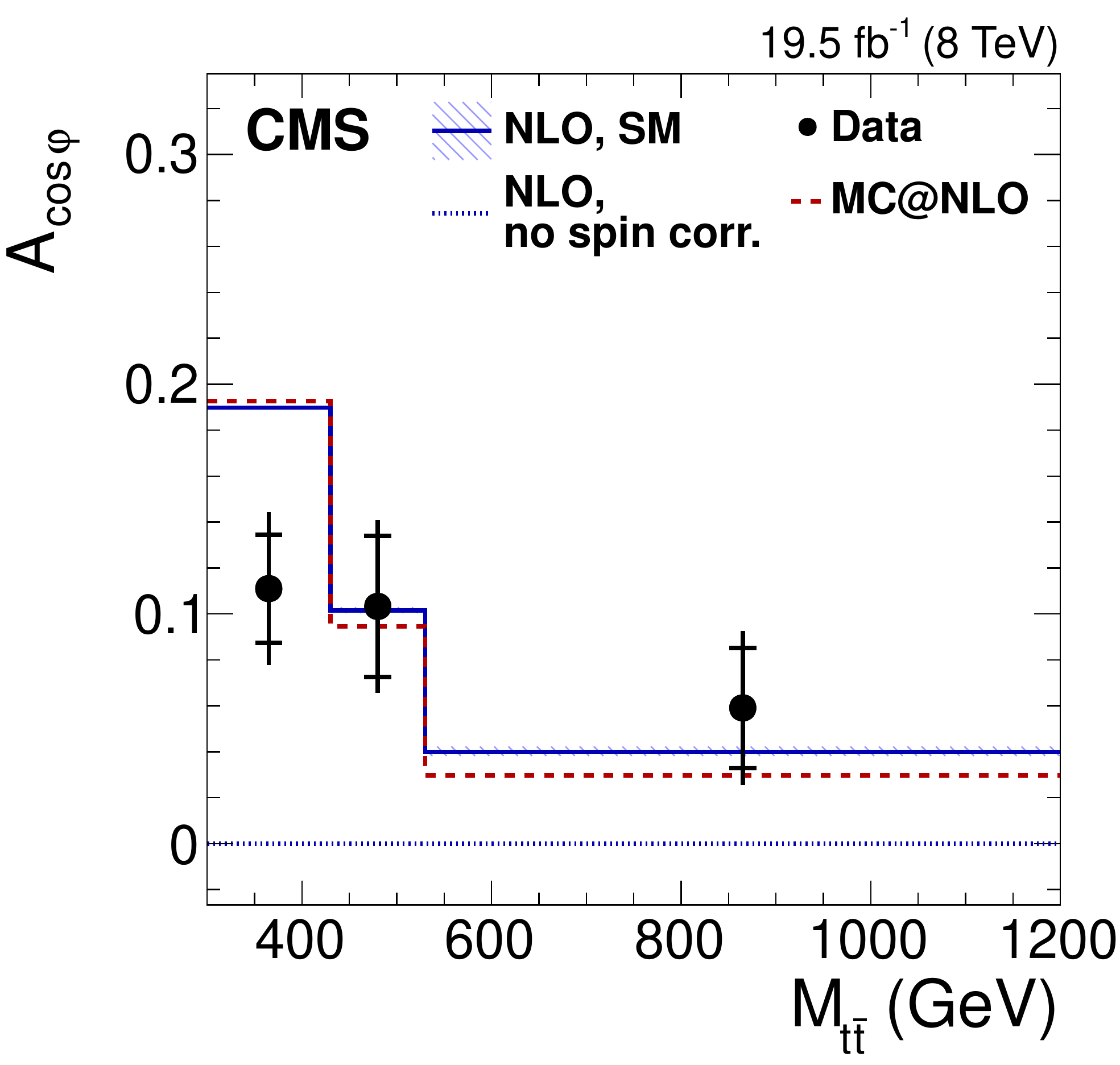}
\includegraphics[width=0.300\linewidth]{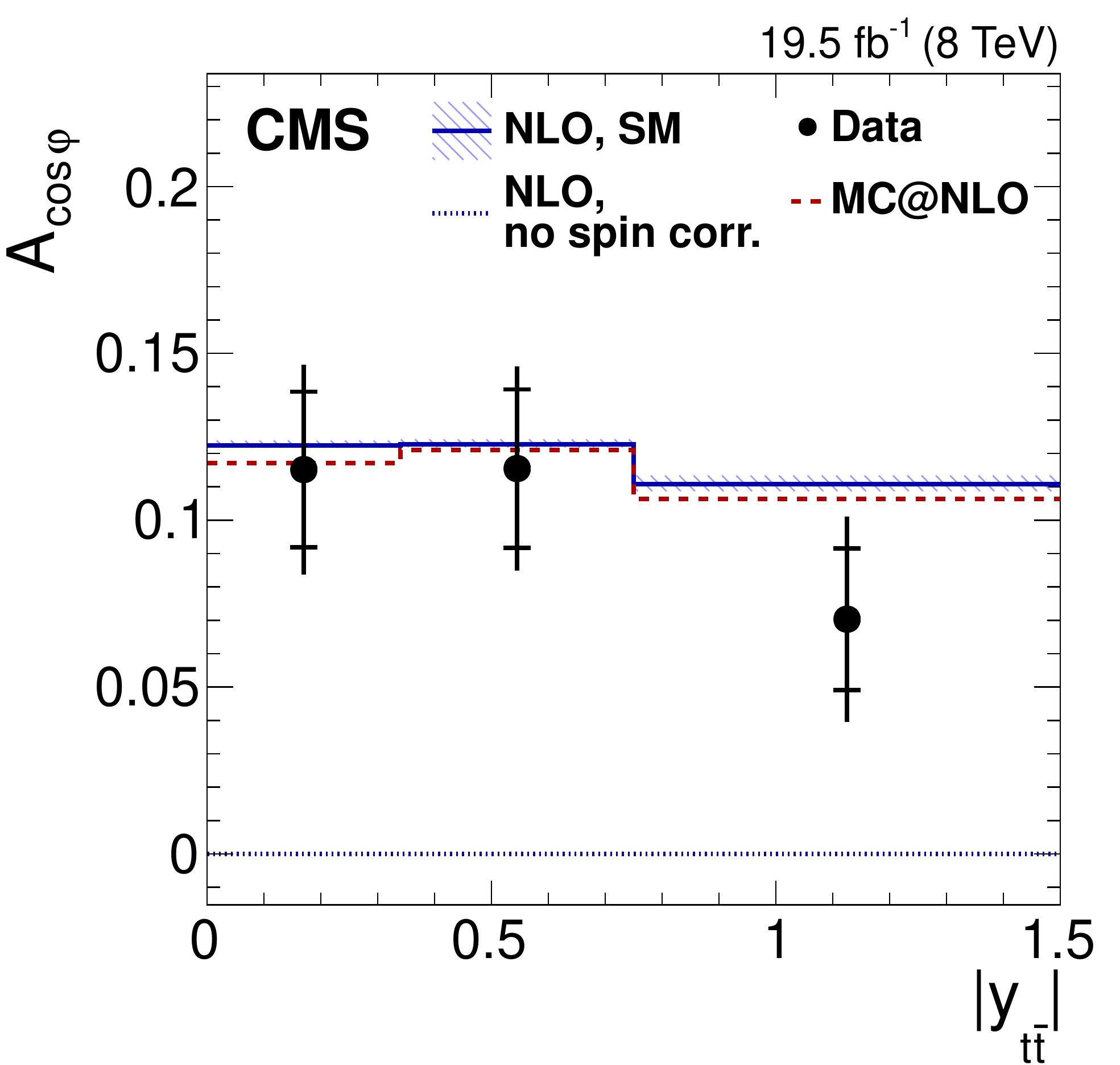}
\includegraphics[width=0.300\linewidth]{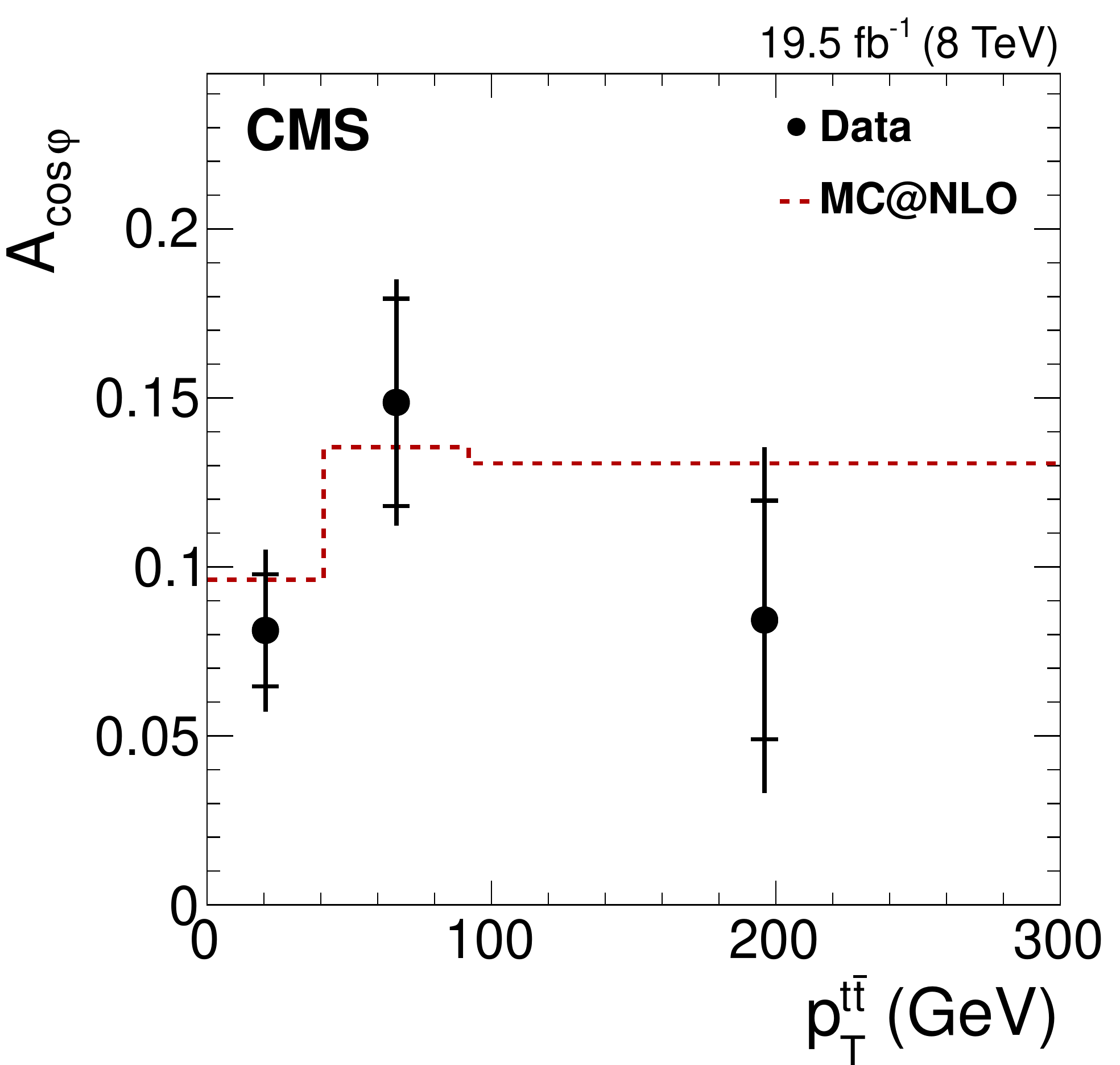}\\
\includegraphics[width=0.300\linewidth]{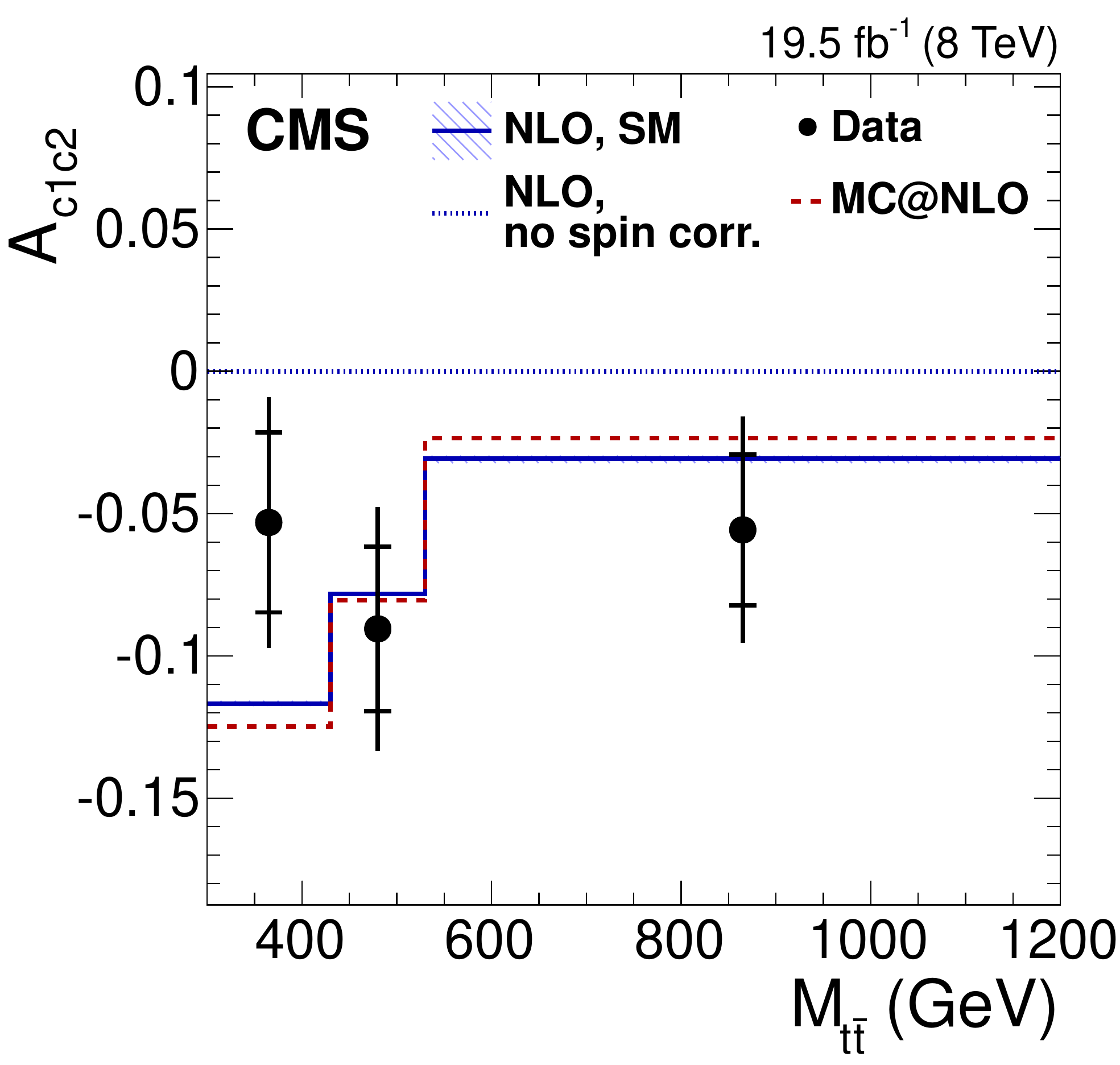}
\includegraphics[width=0.300\linewidth]{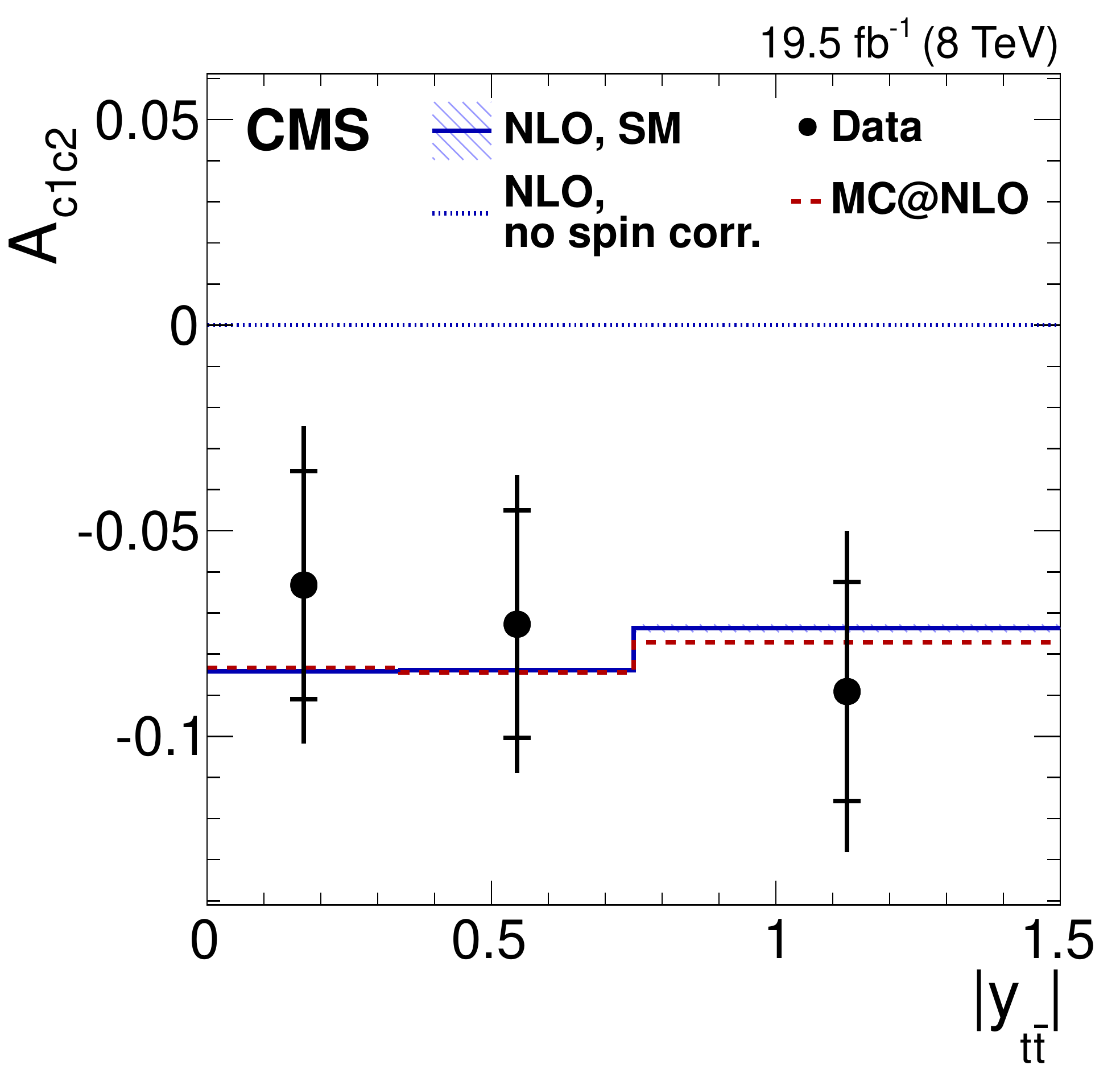}
\includegraphics[width=0.300\linewidth]{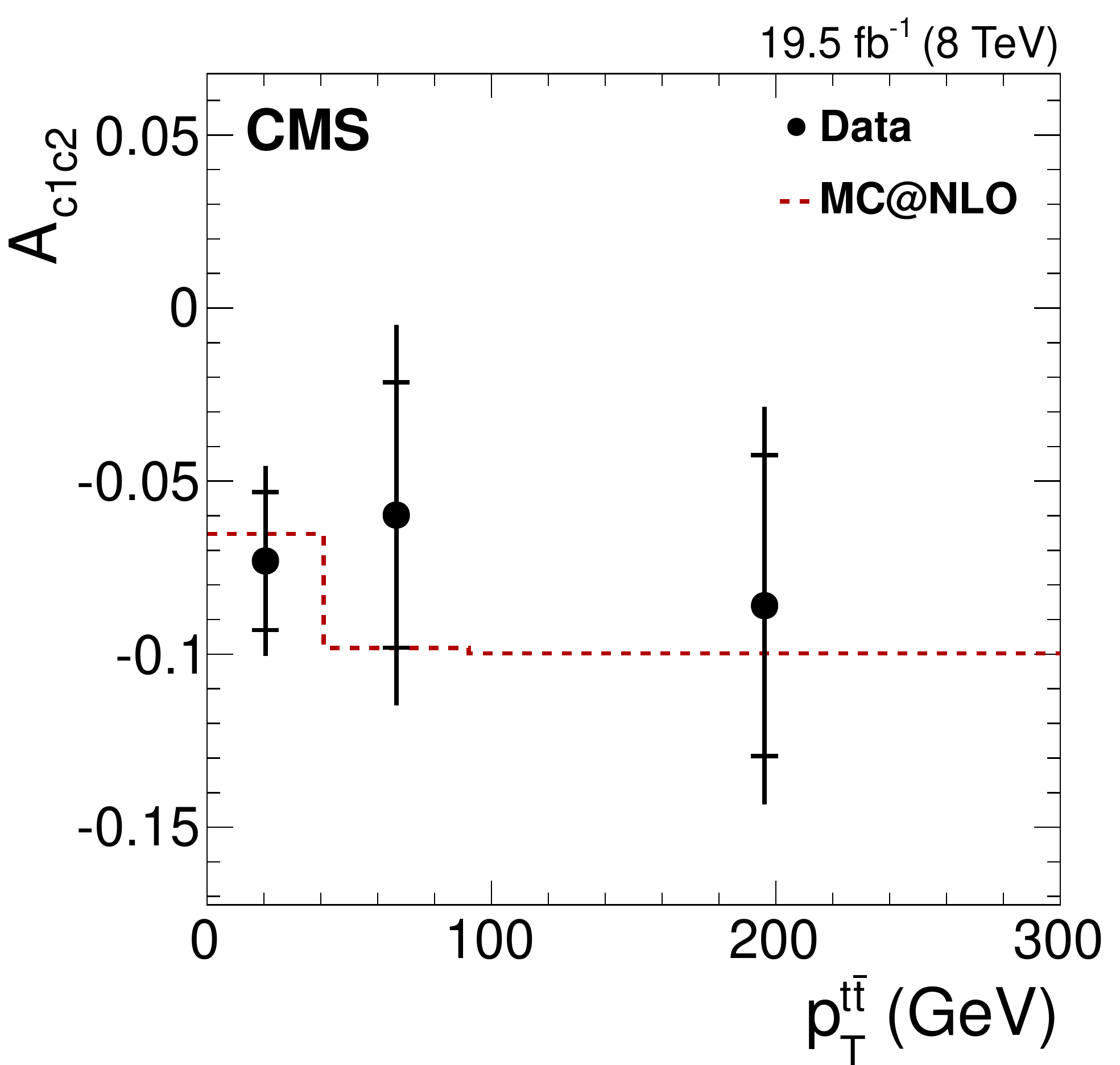}\\
\includegraphics[width=0.300\linewidth]{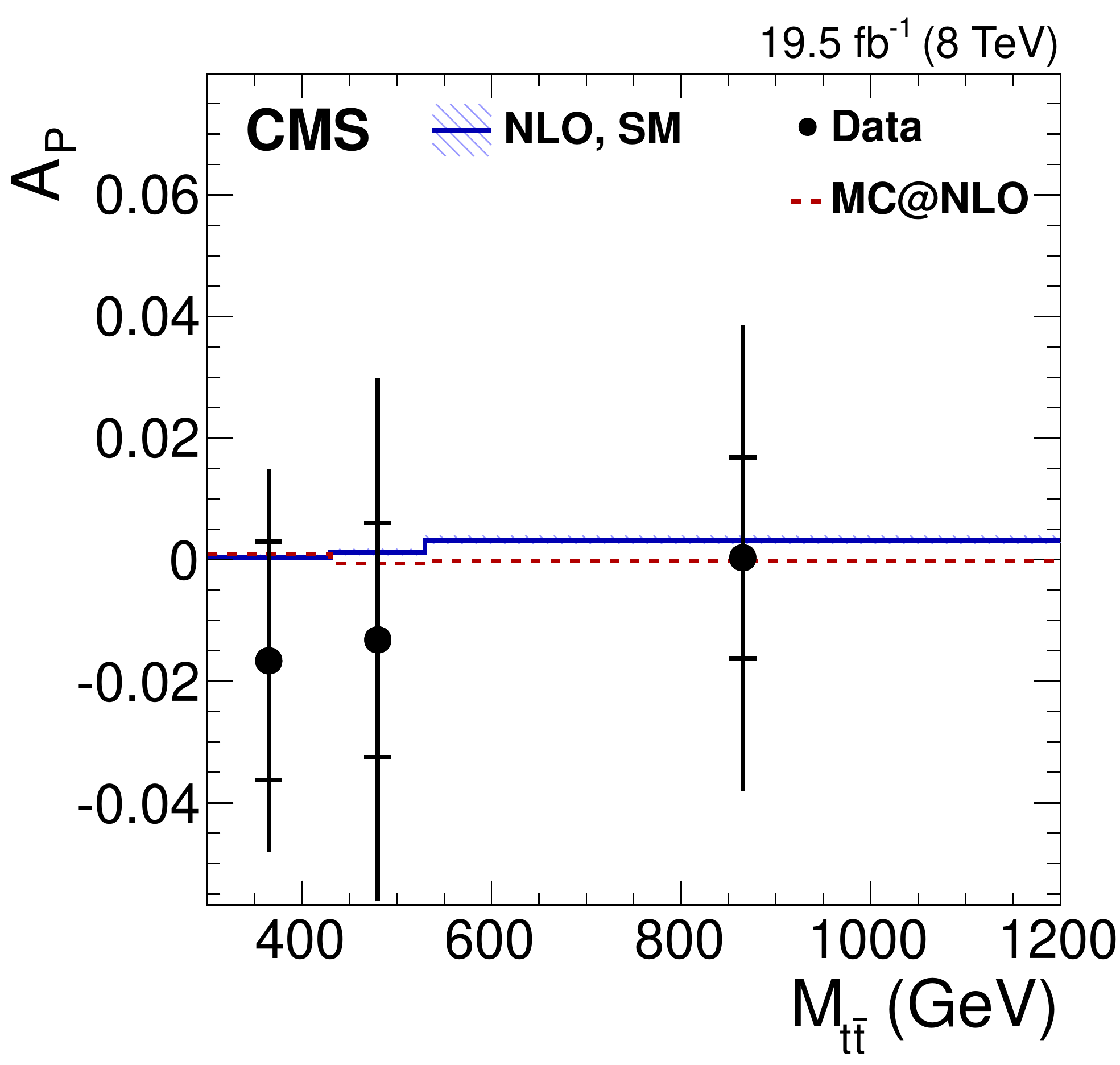}
\includegraphics[width=0.300\linewidth]{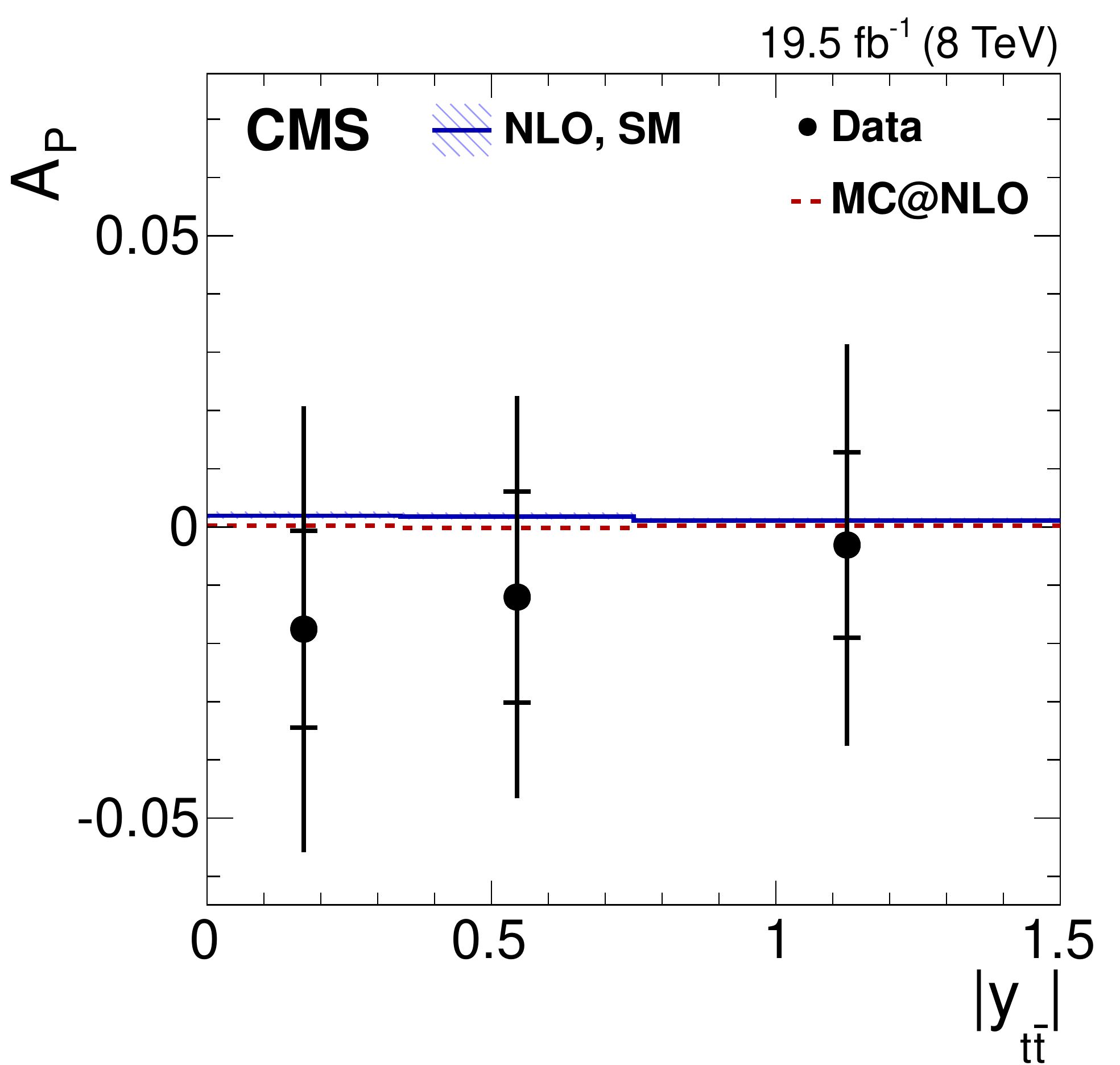}
\includegraphics[width=0.300\linewidth]{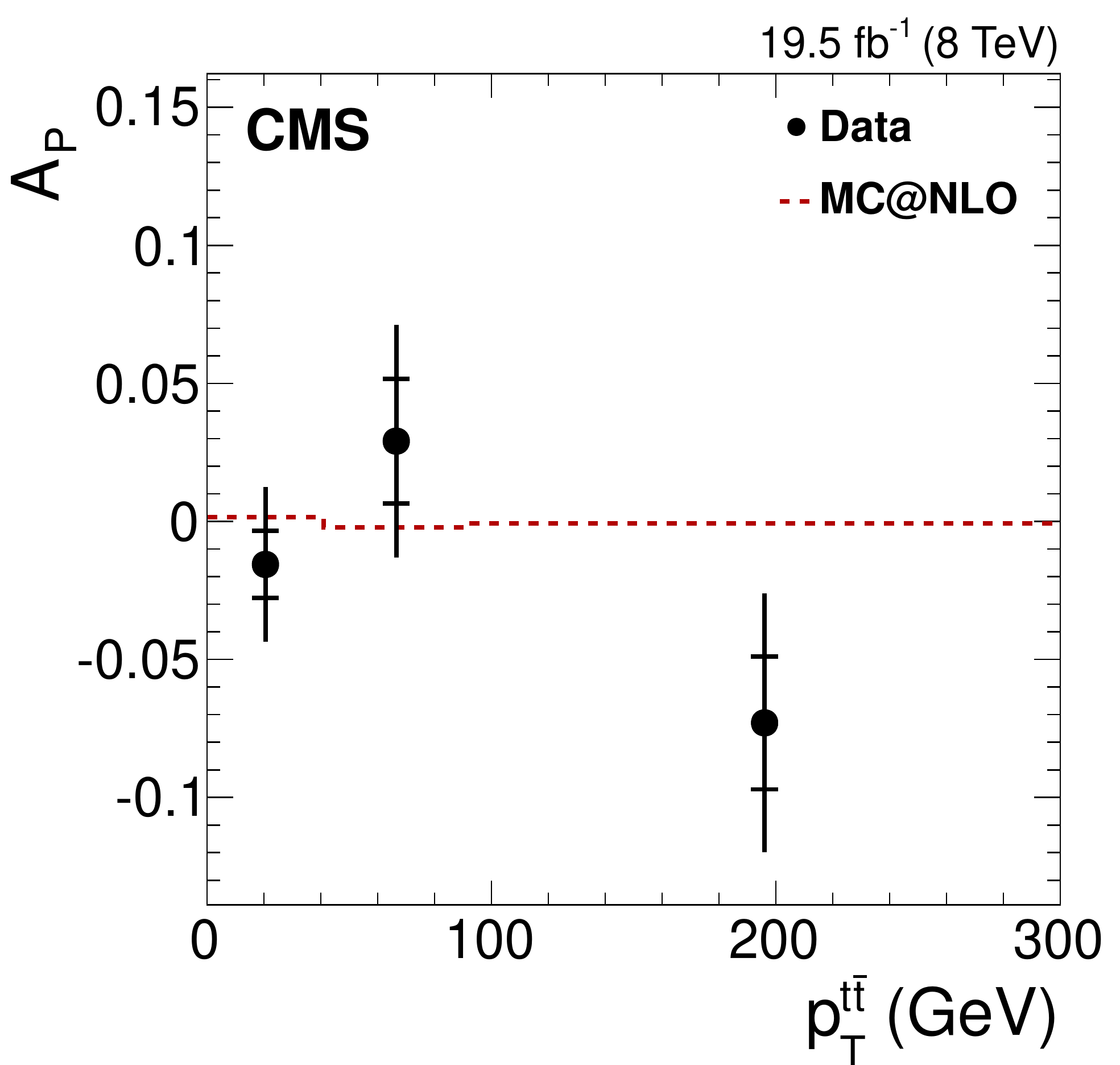}
\caption{\label{fig:AlepCdiff}\protect
Dependence of the four asymmetry variables from data (points) on $M_{\ttbar}$ (left), $\abs{y_{\ttbar}}$ (middle), and $\ptttbar$ (right), obtained from the unfolded double-differential distributions; parton-level predictions from \MCATNLO\ (dashed histograms); and theoretical predictions at \NLO~\cite{theoretical,Bernreuther2013115} with (SM) and without (no spin corr.) spin correlations (solid and dotted histograms, respectively).
The inner and outer vertical bars on the data points represent the statistical and total uncertainties, respectively.
The hatched bands represent variations of $\mu_\mathrm{R}$ and $\mu_\mathrm{F}$ simultaneously up and down by a factor of 2.
The last bin of each plot includes overflow events.
}
\end{figure*}

The dependence of each asymmetry on $M_{\ttbar}$, $\abs{y_{\ttbar}}$, and $\ptttbar$ is extracted from the measured normalized double-differential cross section, and the
results are shown in \fig~\ref{fig:AlepCdiff}.
The measurements are all consistent with the \MCATNLO\ predictions, and with the SM \NLO prediction for the $M_{\ttbar}$ and $\abs{y_{\ttbar}}$ dependencies.
No comparison is made with the \NLO prediction for the $\ptttbar$ dependence because the substantial effect of the parton shower on the $\ptttbar$ distribution means fixed-order NLO calculations are not a sufficiently good approximation of the data.

Compared to the measurement of $A_{\Delta\phi}$ in \tab~\ref{tab:ResultsUnfolded}, the differential measurement in bins of $M_{\ttbar}$ (\fig~\ref{fig:AlepCdiff}, top row, left plot) has a significantly reduced (factor of 2.3) systematic uncertainty associated with the top quark \pt modeling. When the acceptance correction is binned in a variable that is correlated with the top quark \pt (e.g.,\,$M_{\ttbar}$), the top quark \pt reweighting affects the numerator and denominator in the acceptance ratio similarly, leading to a reduction in the associated systematic uncertainty.
The inclusive asymmetry measured from the projection in $\abs{\Delta \phi_{\ell^+\ell^-}}$ of the normalized double-differential cross section
is $A_{\Delta\phi} = 0.095 \pm 0.006\stat \pm 0.007 \syst $, which is converted into the value of $f_{\mathrm{SM}} = 1.12\:^{+\:0.12}_{-\:0.15}$ given in \tab~\ref{tab:fSM}.

\subsection{Limits on new physics}
\label{sec:BSM}

Anomalous $\ttbar \cPg$ couplings can lead to a significant modification of the polarization and spin correlations in \ttbar\ events. A model-independent search can be performed using an effective model of
chromo-magnetic and chromo-electric dipole moments (denoted CMDM and CEDM, respectively).
This study follows the proposal in \reference~\cite{Bernreuther2013115}.
For an anomalous $\ttbar \cPg$ interaction arising from heavy-particle exchange characterized by a mass scale $M \gtrsim m_{\cPqt}$, one can write an effective Lagrangian as

\begin{equation}
 \mathcal{L}_{\mathrm{eff}} = -\frac{\widetilde{\mu}_{\cPqt}}{2} \PAQt \sigma^{\mu\nu} T^a \PQt G^a_{\mu\nu}-\frac{\widetilde{d}_{\cPqt}}{2} \PAQt i\sigma^{\mu\nu} \gamma_5  T^a \PQt G^a_{\mu\nu}, \label{equ:lagr}
\end{equation}

where $\widetilde{\mu}_{\cPqt}$ and $\widetilde{d}_{\cPqt}$ are the CMDM (CP-conserving) and CEDM (CP-violating) dipole moments, $G^a_{\mu\nu}$ is the gluon field strength, and $T^a$ are the QCD fundamental generators. It is usually preferred to define  dimensionless parameters
\begin{equation}
\hat{\mu}_{\cPqt} \equiv \frac{m_{\cPqt}}{g_s} \widetilde{\mu}_{\cPqt},   \;\;\;\;   \hat{d}_{\cPqt} \equiv \frac{m_{\cPqt}}{g_s} \widetilde{d}_{\cPqt},
\end{equation}
where $g_s$ is the QCD coupling constant~\cite{Bernreuther2013115}.
The parameters $\hat{\mu}_{\cPqt}$ and $\hat{d}_{\cPqt}$ correspond to the form factors in the timelike kinematic domain and are therefore complex quantities, here assumed to be constant.
In general, both the real and imaginary parts of $\hat{\mu}_{\cPqt}$ and $\hat{d}_{\cPqt}$ can be determined, but the spin correlations and polarization measured in this paper are only sensitive to $\relmu$ and $\imd$, respectively~\cite{Bernreuther2013115}.

We begin with the determination of $\relmu$ using the measured normalized differential cross section $(1/\sigma)(\rd\sigma/\rd \abs{\Delta \phi_{\ell^+\ell^-}})$.
In the presence of a small new physics (NP) contribution such that $\relmu \ll 1$, one can linearly expand the normalized differential cross section as~\cite{Bernreuther2013115}

\begin{equation}
 \ifthenelse{\boolean{cms@external}}
 {
 \begin{split}
 \frac{1}{\sigma}\frac{\rd\sigma}{\rd\abs{\Delta \phi_{\ell^+\ell^-}}} =  & \left(\frac{1}{\sigma}\frac{\rd\sigma}{\rd\abs{\Delta \phi_{\ell^+\ell^-}}}\right)_{\mathrm{SM}} \\
	& +\ \relmu \, \left(\frac{1}{\sigma}\frac{\rd\sigma}{\rd\abs{\Delta \phi_{\ell^+\ell^-}}}\right)_{\mathrm{NP}}.
\end{split}
 }{
\frac{1}{\sigma}\frac{\rd\sigma}{\rd\abs{\Delta \phi_{\ell^+\ell^-}}} = \left(\frac{1}{\sigma}\frac{\rd\sigma}{\rd\abs{\Delta \phi_{\ell^+\ell^-}}}\right)_{\mathrm{SM}}
	+ \relmu \, \left(\frac{1}{\sigma}\frac{\rd\sigma}{\rd\abs{\Delta \phi_{\ell^+\ell^-}}}\right)_{\mathrm{NP}}.
}	 \label{equ:diff_form}
\end{equation}

\begin{figure}[htb]
\centering
     \includegraphics[width=0.450\textwidth]{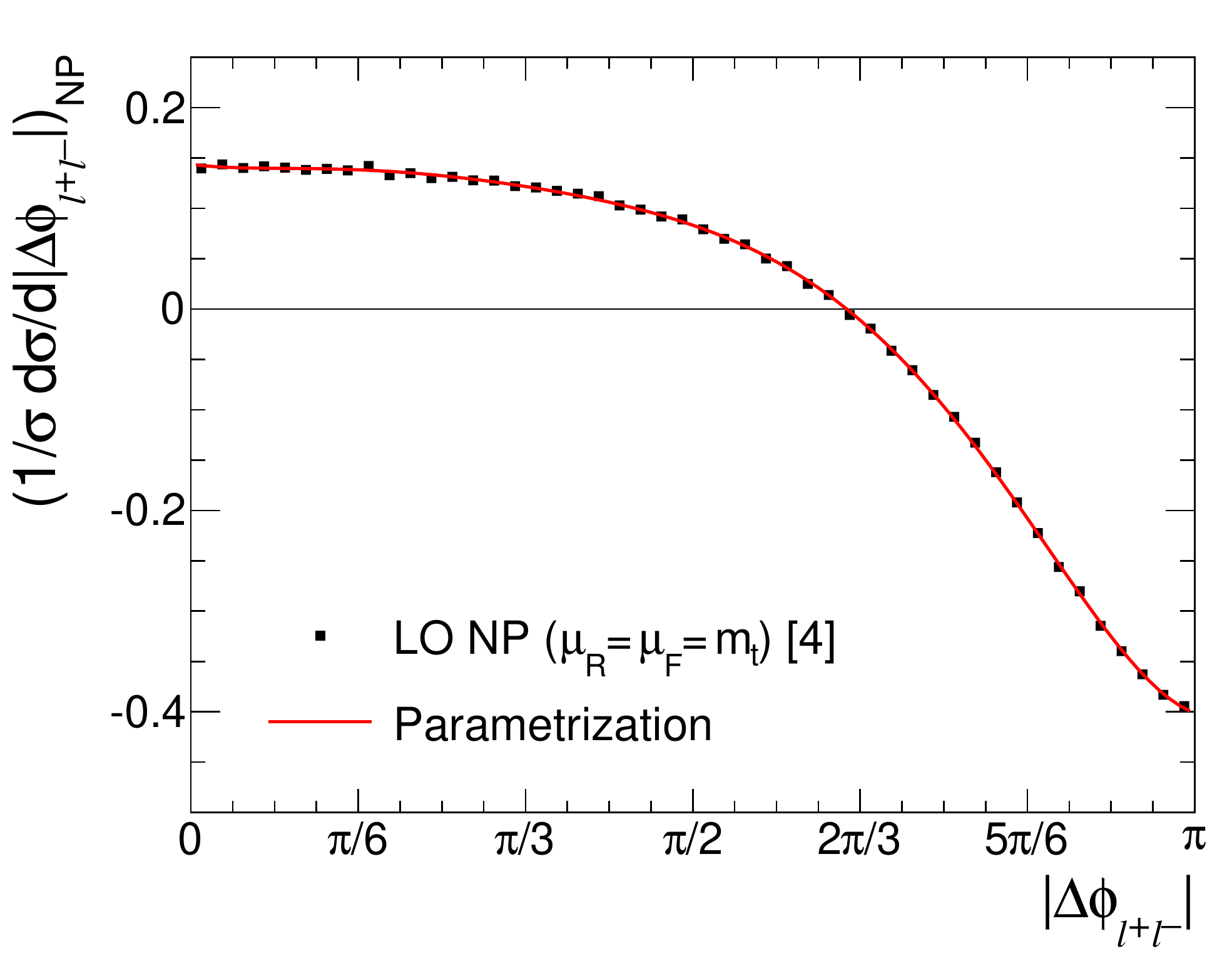}
      \includegraphics[width=0.450\textwidth]{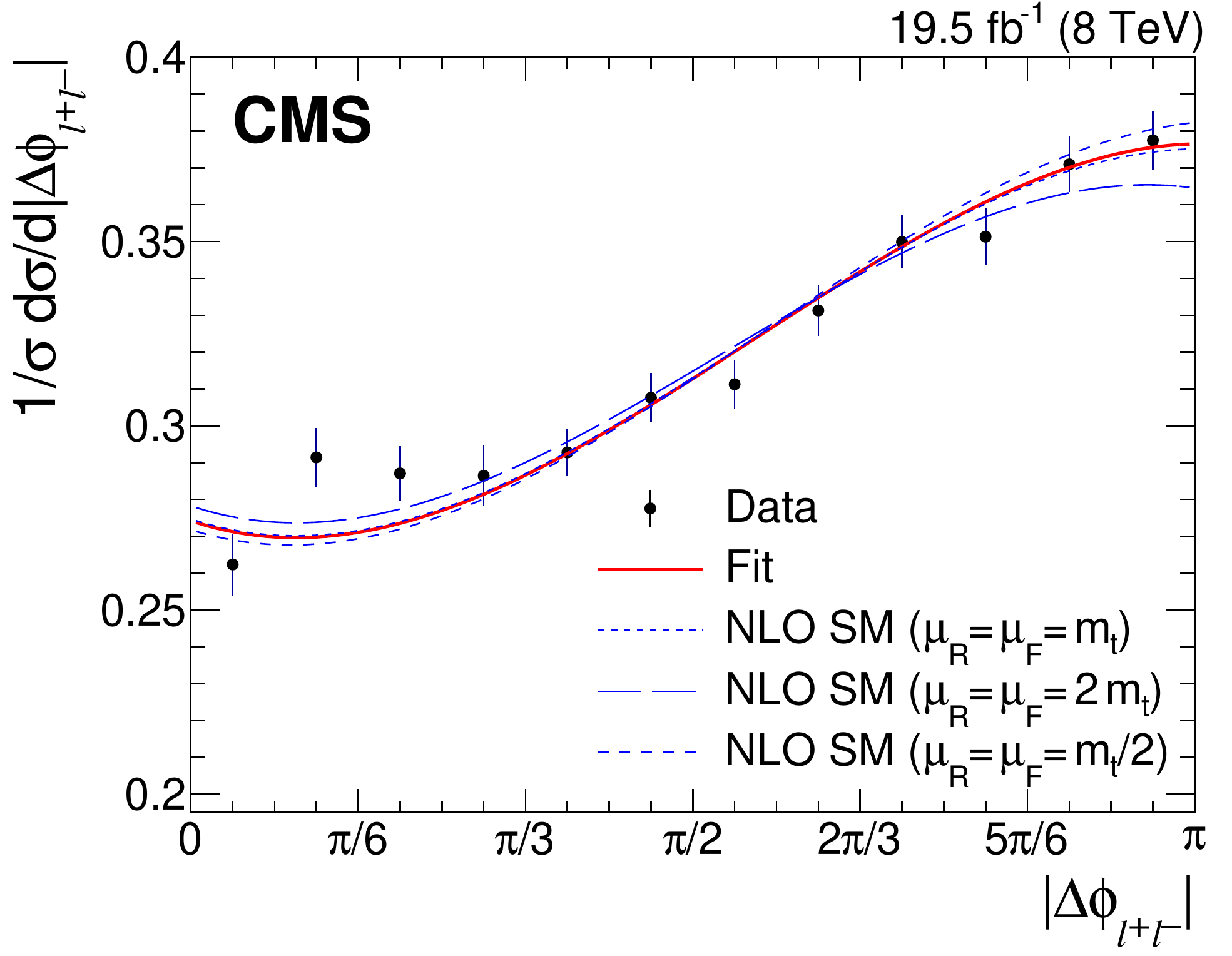}
  \caption{
  \cmsLeft: theoretical prediction from \reference~\cite{Bernreuther2013115} (points) and polynomial parametrization (line) for the contribution from new physics with a nonzero CMDM to
the normalized differential cross section $(1/\sigma)(\rd\sigma/\rd\abs{\Delta \phi_{\ell^+\ell^-}})$,
 for $\relmu \ll 1$.
 \cmsRight: normalized differential cross section from data (points). The solid line corresponds to the result of the fit to the form given in \eqn~(\ref{equ:diff_form}), and the dashed lines show the parametrized SM \NLO predictions for $\mu_\mathrm{R}$ and $\mu_\mathrm{F}$ equal to $m_{\cPqt}$, $2m_{\cPqt}$, and $m_{\cPqt}/2$.
  The vertical bars on the data points represent the total uncertainties.}
  \label{fig:datadistrib_fit_statPlusSyst}
\end{figure}

The predicted shapes of the SM and NP terms in \eqn~(\ref{equ:diff_form}) are shown in \fig~\ref{fig:datadistrib_fit_statPlusSyst}. The NP term arises from interference with SM \ttbar\ production, and therefore gives both positive and negative contributions to the differential cross section.

To measure $\relmu$, the SM and NP contributions to \eqn~(\ref{equ:diff_form}) are parametrized by polynomial functions (shown in \fig~\ref{fig:datadistrib_fit_statPlusSyst}), which are then used in a template fit to the measured normalized differential cross section.
We use the projection in \dphi\ of the measured normalized double-differential cross section in bins of $M_{\ttbar}$ to minimize the systematic uncertainty from top quark \pt\ modeling, as for the extraction of $f_{\mathrm{SM}}$.
The measurement is made under the assumption that the $A$ matrix is unchanged by the presence of NP. Studies of the effects of our selection criteria at the parton level show that this leads to conservative estimates of the experimental uncertainties.
The fit is performed using a $\chi^2$ minimization, accounting for both statistical and systematic uncertainties and their correlations, with $\relmu$ as the only free parameter. The systematic uncertainty arising from the choice of $\mu_\mathrm{R}$ and $\mu_\mathrm{F}$ in the theoretical calculations from \reference~\cite{Bernreuther2013115} is estimated by repeating the fit after varying both scales together up and down by a factor of 2.
This constitutes the dominant source of uncertainty. The proper behavior of the fit is verified using pseudoexperiments.
The result of the fit is $\relmu = -0.006 \pm 0.024$ and is shown graphically in \fig~\ref{fig:datadistrib_fit_statPlusSyst}.
The corresponding 95\% confidence level (CL) interval is $-0.053< \relmu < 0.042$.

The spin correlation coefficient $D$ is also sensitive to $\relmu$, and the CP-violating component of the top quark polarization $P^{\mathrm{CPV}}$ is sensitive to $\imd$. Studies of the effects of our selection criteria at the parton level show that the presence of anomalous top quark chromo moments has no significant effect on the $A$ matrix for either of these variables, and we use this assumption in the derivation of limits on $\relmu$ and $\imd$.

For the $D$ coefficient, \eqn~(\ref{equ:diff_form}) simplifies to $D = D_{\mathrm{SM}} + \relmu \, D_{\mathrm{NP}}$~\cite{Bernreuther2013115}. Using the values from \tab~\ref{tab:ResultsUnfolded}, the relationship $D = -2 A_{\cos\varphi}$, and taking $D_{\mathrm{NP}}=-1.712\pm0.019$ from \reference~\cite{Bernreuther2013115}, we find $\relmu = -0.014 \pm 0.020$, with the corresponding 95\% CL interval $-0.053< \relmu < 0.026$.
The constraints on $\relmu$ from $D$ are stronger than those from the \dphi\ fit because the smaller theoretical uncertainty in the SM \NLO calculation of $D$ compared to that in the \dphi\ distribution outweighs the slightly larger experimental uncertainty.

Similarly, $P^{\mathrm{CPV}}$ is related to $\imd$ via $P^{\mathrm{CPV}} = \imd \, P^{\mathrm{CPV}}_{\mathrm{NP}}$, with $P^{\mathrm{CPV}}_{\mathrm{NP}}=0.482\pm0.003$~\cite{Bernreuther2013115}. We find $\imd = -0.001\pm0.034$, with the corresponding 95\% CL interval $-0.068< \imd < 0.067$.

The $\abs{\Delta \phi_{\ell^+\ell^-}}$ distribution is potentially sensitive to pair-produced scalar top quark partners (top squarks) that decay to produce a top quark and antiquark with no additional visible particles.
The spin-zero particles transmit no spin information from the initial state to the final-state top quarks, meaning such events look much like uncorrelated \ttbar\ events.
We assess the sensitivity of the
measured $\abs{\Delta \phi_{\ell^+\ell^-}}$ distribution to pair-produced top squarks with mass equal to $m_{\cPqt}$.
As seen from the measurement of $f_{\mathrm{SM}}$ in the last row of \tab~\ref{tab:fSM}, the dominant source of uncertainty is the theoretical scale uncertainty in the $\abs{\Delta \phi_{\ell^+\ell^-}}$ distribution.
The result is that no exclusion limits on top squarks can be set using the $\abs{\Delta \phi_{\ell^+\ell^-}}$ normalized differential cross section alone, and the additional sensitivity it would bring in combination with the inclusive
measurement of the cross section would be marginal.

\section{Summary}

Measurements of the \ttbar\ spin correlations and the top quark polarization have been presented in the \ttbar\ dilepton final states (\eepm, \empm, and \mmpm),
using angular distributions unfolded to the parton level and as a function of the \ttbar -system variables $M_{\ttbar}$, $\abs{y_{\ttbar}}$, and $\ptttbar$.
The data sample corresponds to an integrated luminosity of 19.5\fbinv from pp collisions at $\sqrt{s}=8$\TeV, collected by the CMS experiment at the LHC.

For the spin correlation coefficients, we measure $C_{\mathrm{hel}} = 0.278 \pm 0.084$ and $D = 0.205 \pm 0.031$. The measurements sensitive to spin correlations are translated into determinations of $f_{\mathrm{SM}}$, the strength of the spin correlations relative to the SM prediction. The most precise result comes from the measurement of $A_{\Delta\phi} = 0.095 \pm 0.006\:\mathrm{(stat)} \pm 0.007 \:\mathrm{(syst)}$, yielding $f_{\mathrm{SM}} = 1.12\:^{+\:0.12}_{-\:0.15}$.
The SM (CP-conserving) top quark polarization is measured to be $P = -0.022 \pm 0.058$,
while the CP-violating component is found to be $P^{\mathrm{CPV}} = 0.000 \pm 0.016$.
All measurements are in agreement with the SM expectations, and help
constrain theories of physics beyond the SM.

The measured top quark spin observables are compared to theoretical predictions in order to search for hypothetical top quark anomalous couplings.
No evidence of new physics is observed, and exclusion limits on the real part of the chromo-magnetic dipole moment $\relmu$ and the imaginary part of the chromo-electric dipole moment $\imd$ are evaluated.
Values outside the intervals $-0.053<\relmu<0.026$
and $-0.068< \imd < 0.067$
are excluded at the 95\% confidence level,
the first such measurements to date.

\begin{acknowledgments}

\hyphenation{Bundes-ministerium Forschungs-gemeinschaft Forschungs-zentren}

We would like to thank W.~Bernreuther and Z.-G.~Si for calculating the theoretical predictions for this paper, and for studies of the effect of anomalous top quark chromo moments on the acceptance of our selection criteria at the parton level.
We congratulate our colleagues in the CERN accelerator departments for the excellent performance of the LHC and thank the technical and administrative staffs at CERN and at other CMS institutes for their contributions to the success of the CMS effort. In addition, we gratefully acknowledge the computing centers and personnel of the Worldwide LHC Computing Grid for delivering so effectively the computing infrastructure essential to our analyses. Finally, we acknowledge the enduring support for the construction and operation of the LHC and the CMS detector provided by the following funding agencies: the Austrian Federal Ministry of Science, Research and Economy and the Austrian Science Fund; the Belgian Fonds de la Recherche Scientifique, and Fonds voor Wetenschappelijk Onderzoek; the Brazilian Funding Agencies (CNPq, CAPES, FAPERJ, and FAPESP); the Bulgarian Ministry of Education and Science; CERN; the Chinese Academy of Sciences, Ministry of Science and Technology, and National Natural Science Foundation of China; the Colombian Funding Agency (COLCIENCIAS); the Croatian Ministry of Science, Education and Sport, and the Croatian Science Foundation; the Research Promotion Foundation, Cyprus; the Ministry of Education and Research, Estonian Research Council via IUT23-4 and IUT23-6 and European Regional Development Fund, Estonia; the Academy of Finland, Finnish Ministry of Education and Culture, and Helsinki Institute of Physics; the Institut National de Physique Nucl\'eaire et de Physique des Particules~/~CNRS, and Commissariat \`a l'\'Energie Atomique et aux \'Energies Alternatives~/~CEA, France; the Bundesministerium f\"ur Bildung und Forschung, Deutsche Forschungsgemeinschaft, and Helmholtz-Gemeinschaft Deutscher Forschungszentren, Germany; the General Secretariat for Research and Technology, Greece; the National Scientific Research Foundation, and National Innovation Office, Hungary; the Department of Atomic Energy and the Department of Science and Technology, India; the Institute for Studies in Theoretical Physics and Mathematics, Iran; the Science Foundation, Ireland; the Istituto Nazionale di Fisica Nucleare, Italy; the Ministry of Science, ICT and Future Planning, and National Research Foundation (NRF), Republic of Korea; the Lithuanian Academy of Sciences; the Ministry of Education, and University of Malaya (Malaysia); the Mexican Funding Agencies (CINVESTAV, CONACYT, SEP, and UASLP-FAI); the Ministry of Business, Innovation and Employment, New Zealand; the Pakistan Atomic Energy Commission; the Ministry of Science and Higher Education and the National Science Centre, Poland; the Funda\c{c}\~ao para a Ci\^encia e a Tecnologia, Portugal; JINR, Dubna; the Ministry of Education and Science of the Russian Federation, the Federal Agency of Atomic Energy of the Russian Federation, Russian Academy of Sciences, and the Russian Foundation for Basic Research; the Ministry of Education, Science and Technological Development of Serbia; the Secretar\'{\i}a de Estado de Investigaci\'on, Desarrollo e Innovaci\'on and Programa Consolider-Ingenio 2010, Spain; the Swiss Funding Agencies (ETH Board, ETH Zurich, PSI, SNF, UniZH, Canton Zurich, and SER); the Ministry of Science and Technology, Taipei; the Thailand Center of Excellence in Physics, the Institute for the Promotion of Teaching Science and Technology of Thailand, Special Task Force for Activating Research and the National Science and Technology Development Agency of Thailand; the Scientific and Technical Research Council of Turkey, and Turkish Atomic Energy Authority; the National Academy of Sciences of Ukraine, and State Fund for Fundamental Researches, Ukraine; the Science and Technology Facilities Council, UK; the US Department of Energy, and the US National Science Foundation.

Individuals have received support from the Marie-Curie programme and the European Research Council and EPLANET (European Union); the Leventis Foundation; the A. P. Sloan Foundation; the Alexander von Humboldt Foundation; the Belgian Federal Science Policy Office; the Fonds pour la Formation \`a la Recherche dans l'Industrie et dans l'Agriculture (FRIA-Belgium); the Agentschap voor Innovatie door Wetenschap en Technologie (IWT-Belgium); the Ministry of Education, Youth and Sports (MEYS) of the Czech Republic; the Council of Science and Industrial Research, India; the HOMING PLUS programme of the Foundation for Polish Science, cofinanced from European Union, Regional Development Fund; the OPUS programme of the National Science Center (Poland); the Compagnia di San Paolo (Torino); MIUR project 20108T4XTM (Italy); the Thalis and Aristeia programmes cofinanced by EU-ESF and the Greek NSRF; the National Priorities Research Program by Qatar National Research Fund; the Rachadapisek Sompot Fund for Postdoctoral Fellowship, Chulalongkorn University (Thailand); the Chulalongkorn Academic into Its 2nd Century Project Advancement Project (Thailand); and the Welch Foundation, contract C-1845.

\end{acknowledgments}

\clearpage

\bibliography{auto_generated}

\cleardoublepage \appendix\section{The CMS Collaboration \label{app:collab}}\begin{sloppypar}\hyphenpenalty=5000\widowpenalty=500\clubpenalty=5000\textbf{Yerevan Physics Institute,  Yerevan,  Armenia}\\*[0pt]
V.~Khachatryan, A.M.~Sirunyan, A.~Tumasyan
\vskip\cmsinstskip
\textbf{Institut f\"{u}r Hochenergiephysik der OeAW,  Wien,  Austria}\\*[0pt]
W.~Adam, E.~Asilar, T.~Bergauer, J.~Brandstetter, E.~Brondolin, M.~Dragicevic, J.~Er\"{o}, M.~Flechl, M.~Friedl, R.~Fr\"{u}hwirth\cmsAuthorMark{1}, V.M.~Ghete, C.~Hartl, N.~H\"{o}rmann, J.~Hrubec, M.~Jeitler\cmsAuthorMark{1}, V.~Kn\"{u}nz, A.~K\"{o}nig, M.~Krammer\cmsAuthorMark{1}, I.~Kr\"{a}tschmer, D.~Liko, T.~Matsushita, I.~Mikulec, D.~Rabady\cmsAuthorMark{2}, N.~Rad, B.~Rahbaran, H.~Rohringer, J.~Schieck\cmsAuthorMark{1}, R.~Sch\"{o}fbeck, J.~Strauss, W.~Treberer-Treberspurg, W.~Waltenberger, C.-E.~Wulz\cmsAuthorMark{1}
% ...
\vskip\cmsinstskip
% ...
1:~~Also at Vienna University of Technology, Vienna, Austria\\
2:~~Also at CERN, European Organization for Nuclear Research, Geneva, Switzerland\\
% ...

\end{sloppypar}
\end{document}